\documentclass[11pt]{article}
\usepackage{epsfig}
\usepackage{amssymb,amsmath}
\usepackage{epigraph}

\newcommand{\beq}{\begin{equation}}
\newcommand{\eeq}{\end{equation}}
\newcommand{\bea}{\begin{eqnarray}}
\newcommand{\eea}{\end{eqnarray}}

\begin{document}



\begin{center}

{\LARGE
QLBS: Q-Learner in the Black-Scholes(-Merton) Worlds    
}
\vskip1.0cm
{\Large Igor Halperin} \\
\vskip0.5cm
NYU Tandon School of Engineering \\
\vskip0.5cm
{\small e-mail: $igor.halperin@nyu.edu $}
\vskip0.5cm
First version: December 17, 2017 \\
This version: \today \\

\vskip1.0cm
{\Large Abstract:\\}
\end{center}
\parbox[t]{\textwidth}{
This paper presents a discrete-time option pricing model that is rooted in Reinforcement Learning (RL), and more specifically in the famous Q-Learning method of RL. We construct a risk-adjusted Markov Decision Process for a discrete-time version of the classical Black-Scholes-Merton (BSM) model, where the option price is an optimal Q-function, while the optimal hedge is a second argument of this optimal Q-function, so that both the price and hedge are parts of the same formula. Pricing is done by learning to dynamically optimize risk-adjusted returns for an option replicating portfolio, as in the Markowitz portfolio theory.  
Using Q-Learning and related methods, once created in a parametric setting, the model is able to go model-free and learn to price and hedge an option directly from data, and without an explicit model of the world. 
This suggests that RL may provide efficient data-driven and model-free methods for optimal pricing and hedging of options, once we depart from the academic continuous-time limit, and vice versa, option pricing methods developed in Mathematical Finance may be viewed as special cases of model-based Reinforcement Learning. Further, due to simplicity and tractability of our model  which only needs basic linear algebra (plus Monte Carlo simulation, if we work with synthetic data), and its close relation to the original BSM model, we suggest that our model could be used for benchmarking of different RL algorithms for financial trading applications.     

 }

 \newcounter{helpfootnote}
\setcounter{helpfootnote}{\thefootnote} 
\renewcommand{\thefootnote}{\fnsymbol{footnote}}
\setcounter{footnote}{0}
\footnotetext{
I would like to thank Eric Berger, Jean-Philippe Bouchaud, Peter Carr, Andrey Itkin, and Alexander Lipton for comments and discussions on the first draft of this manuscript.
All remaining errors are my own.
This work is dedicated to my wife Lola on the occasion of her birthday and receiving a doctoral degree.}     

 \renewcommand{\thefootnote}{\arabic{footnote}}
\setcounter{footnote}{\thehelpfootnote} 

\newpage
 
\section{Introduction}

Reinforcement learning (RL) is one of the most fascinating sub-fields of machine learning (ML) \cite{SB}. While being more than 50 years old, it seems to attract ever-growing interest of ML researchers. In particular, in finance, many tasks in trading or investment decisions seem well suited for RL approaches. At present, literature on the topic is sparse, and moreover scattered among different financial applications areas. 

This paper suggests to try reinforcement learning in a well-understood financial setting that could be thought of as a financial analogue of the classical inverted pole problem, a  veritable test case for generations of reinforcement learning 
models \cite{SB}. This environment and other physics-inspired simulated environments are available, in particular, in the popular RL research platform OpenAI Gym \cite{OpenAIGym}.
Likewise, a similar ``simplest possible, but not a simpler one" setting could be used for financial applications of reinforcement learning as a testing laboratory for exploration and benchmarking of different RL algorithms for financial applications.

 Here we propose such a testing environment for reinforcement learning in finance. As we will show below, it is very flexible and extensible. In particular, it allows a researcher to benchmark both discrete-action and continuous-action reinforcement learning algorithms. On the side of finance, it offers a look into the problems of hedging/trading and pricing in financial markets - all the main elements of many financial tasks, but in a controllable and well understood environment. 
 
 A model presented in this paper may also be of an independent interest as a data-driven derivatives pricing model rooted in RL. While we present a simplest possible setting, it can be extended in many practically interesting ways, including early exercises, multiple factors, transaction costs, and so on.  
  
Our model extends the celebrated Black-Scholes-Merton (BSM) model, also known as the Black-Scholes (BS) 
model, a cornerstone of modern quantitative finance \cite{BS,Merton}. 
The core idea of the BSM model is that options or other financial derivatives can be priced using the relative value approach to asset pricing, which prices assets in terms of other, tradable assets. 
The relative pricing method for options is known as dynamic option replication.
 It is based on the observation that an option payoff depends only on the price of a stock at expiry of the option. Therefore, if we neglect other sources of uncertainly such as stochastic volatility, the option value at arbitrary times before the expiry should only depend on the stock value. 
 
This makes it possible to mimic the option using a simple portfolio made of the underlying stock and cash, which is called the hedge portfolio. 
The hedge portfolio is dynamically managed by continuously rebalancing its wealth between the stock and cash. Moreover, this is done in a self-financing way, meaning that there are no cash inflows/outflows in the portfolio except at the time of inception. The objective of dynamic replication is to mimic the option using the hedge portfolio as closely as possible.


In the continuous-time setting of the original BSM model, it turns out that such dynamic replication can be made exact by a continuous rebalancing of the hedge portfolio between the stock and cash.
This makes the total portfolio made of the option and the hedge portfolio {\it instantaneously risk-free}, or equivalently it makes the option instantaneously perfectly replicable in terms of the stock and cash. 
As risk of mis-hedging between the option and its underlying is instantaneously eliminated, the full portfolio involving the option and its hedge should earn a risk-free rate. The option price in this limit does not depend on risk preferences of investors, because the mis-hedging risk is completely eliminated, leading to 
a unique option price equal to the price of the replicating portfolio. This result of the BSM model is somewhat paradoxical, as it implies  that options are altogether completely {\it redundant}, as they can always be perfectly replicated by a simple portfolio made of a stock and a bond.

%
%

If this were indeed the case in the real life, i.e. options were totally redundant, nobody would ever trade them, except 
possibly for very bored traders. Yet, option trading is a multi-billion business where people make and lose money daily. Traders use options and other financial derivatives both as investment vehicles and as hedging instruments. Options are {\it not} redundant.  


The reason that options are not redundant is that they are {\it not} risk-free, notwithstanding the main proposition of the classical BSM model. Option pricing involve multiple sources of risk including volatility risk, bid-ask spreads, transaction costs etc., see e.g. \cite{Wilmott}. 
In this paper, we focus on mis-hedging risk and leave other sources of risk in options aside. 
The reason is that mis-hedging risk reappears once we abandon a single most unrealistic assumption of the BSM model, namely the assumption of a continuous re-hedging.



Taking this assumption away means adding more realism to a model.
Indeed, in option markets, rebalancing of option replication  portfolio happens at a finite frequency, e.g. daily. A frequent rebalancing can be costly due to transaction costs which are altogether neglected in the classical BSM model.  When transaction costs are added, a formal continuous-time limit may not even exist as it leads to formally infinite option prices due to an infinite number of portfolio rebalancing acts. The frequency of re-hedging specifies a natural time-scale in the problem, therefore we work in discrete time with time steps equal to re-hedging periods.  

With a finite rebalancing frequency, perfect replication is no longer feasible, and the replicating portfolio will in general be different from the option value according to the amount of hedge slippage. The latter depends on the stock price evolution between consecutive re-balancing acts for the portfolio. Respectively, in absence of perfect replication, a hedged option position carries some mis-hedging risk which the option buyer or seller should be compensated for. This means that once we revert from the idealized setting of continuous-time finance to a realistic setting of discrete-time finance, option pricing becomes dependent on investors' risk preferences.

If we take a view of an option seller agent in such a discrete-time setting, its objective  should be to minimize some measure of slippage risk, also referred to as a risk-adjusted cost of hedging the option, by dynamic option replication. 
When viewed over a lifetime of an option, this can be considered a sequential decision-making process of minimization of slippage cost (or equivalently maximization of rewards determined as negative costs). 
While such a discrete-time approach converges to the Black-Scholes formulation in the limit 
of vanishing time steps, it both offers a more realistic setting,  
and  allows one to focus on the key objective of option trading and pricing, which is {\it risk minimization by hedging in a sequential decision-making process}.
This makes option pricing amenable to methods of reinforcement learning, and indeed, as we will see below, option pricing and hedging in discrete time 
{\it amounts} to reinforcement learning. 

Casting option pricing as a reinforcement learning task offers a few interesting insights.
First, if we select a specific model for the stock price dynamics, we can use model-based reinforcement learning as a powerful sample-based (Monte Carlo) computational approach. The latter may be advantageous to other numerical methods such as finite differences for computing option prices and hedge ratios, especially when dimensionality of the state space goes beyond three or four. Second, we may rely on model-free reinforcement learning methods such as Q-learning, and bypass the need to build a model of stock price dynamics altogether.  RL provides a framework for model-free learning of option prices and hedges\footnote{Here we use the notion of model-free learning in the same sense it is normally used in the machine learning literature, namely as a method that does not rely on an explicit model of feature dynamics. Option prices and hedge ratios in our approach depend on a model of rewards, and in this sense are model-dependent.}. While we only consider the simplest setting for a reinforcement learning approach to pricing and hedging of  European vanilla options (e.g. (put or call options), the approach can be extended in straightforward manner to handle more complex instruments including options on multiple assets, early exercises, option portfolios, market frictions etc.

The model presented in this paper is referred to as the QLBS model, 
in recognition of the fact that it combines the Q-learning method of 
Watkins \cite{Watkins_1989, Watkins} with the method of dynamic option replication of the (time-discretized) Black-Scholes model. 
 As Q-learning is a model-free method, this means that the QLBS model is also model-free. More accurately, it is {\it distribution-free}: option prices in this approach depend on the chosen utility function, but do not rely on any model for the stock price distribution, and instead use only samples from this distribution.

The QLBS model may also be of interest as a financial model which relates to the literature on hedging and pricing in incomplete markets \cite{FS, Schweizer, CK, 
PB, Kapoor, Grau}.
Unlike many previous models of this sort,  QLBS ensures a full consistency of hedging and pricing at each time step, all within an efficient and data-driven Q-Learning algorithm. Additionally, it  {\it extends} the discrete-time BSM model. Extending Markowitz portfolio theory \cite{Markowitz} to a multi-period setting, Sect.~\ref{sect:DT_BSM} incorporates a drift in a risk/return analysis of the option's hedge portfolio. This extension allows one to consider both {\it hedging and speculation} with options in a consistent way within the {\it same} model, which is a challenge for the standard BSM model or its 
``phenomenological" generalizations, see e.g. \cite{Wilmott}.  

Following this approach, it turns out that all results of the 
classical BSM model \cite{BS, Merton} can be obtained as a continuous-time limit $ \Delta t \rightarrow 0 $ of a multi-period self-financing version of the Markowitz portfolio theory \cite{Markowitz}, for a special case when the portfolio is made only of a stock and cash, and the stock price dynamics are lognormal. However, this limit is {\it degenerate}: all fluctuations of the ``true" option price asymptotically decay in this limit, resulting in a deterministic option price which is independent of risk preferences of an investor. However, as long as  the time step  $ \Delta t $ is kept finite, both risk of option mis-hedging and dependence of the option price on investor risk preferences persist.



Perhaps a bit surprisingly, we find that keeping the time step $ \Delta t $ finite makes the resulting model in a sense even simpler than the classical continuous time BSM model that produces two different formulas for the option price and its delta (the optimal hedge), both involving transcendent functions like $ \mathcal{N}(d_1) $ or $   \mathcal{N}(d_2) $ , see e.g. \cite{Wilmott}.  In the QLBS approach, both option price and its delta are part of the same object (the optimal Q-function) whose calculation involves only a number of linear regressions, and nothing else, assuming that a historical dataset of market prices and hedge re-balancing trades is available.

The computational framework of the QLBS model can be used for either real data or data generated from a known model. If we choose to enforce a model for the stock price dynamics, the QLBS model formulation can be used a a tool for approximate dynamic programming (ADP) or model-based reinforcement learning.  As mentioned above, such setting can be used to benchmark different RL methods for option pricing. In this paper, we choose the BSM model as a benchmark model, however the same approach could be used for other popular models such as e.g. the GARCH model.   


To the extent that option pricing in discrete time amounts to dynamic programming (aka model-based RL), if a model is known,
we may say that traditional continuous-time option pricing models correspond to a continuous time limit of model-based Reinforcement Learning. 
In such a limit, all data requirements are reduced, for the particular case of the BSM model, to just {\it two} numbers - the current stock price and volatility, because these are the only two statistics that define transition probabilities in this model.  

On the other hand, the QLBS model can also be used with {\it real} data, when the dynamics  are unknown.  
In this case, we can try methods of  
reinforcement learning to learn the optimal hedge and price directly from {\it trading data} for a dynamic hedge portfolio. 
We can apply methods of value-based reinforcement learning \cite{SB} to {\it approximately} solve the Bellman optimality equation without any knowledge of model's dynamics, but relying solely on {\it sample data}. 
While a setting developed below involves only one factor (one stock price), the model is straightforward to extend to multiple factors, different option payoffs, etc.  
Such possible extensions, which would add computational complexity but leave the conceptual framework intact, are left here for a future work.

The paper is organized as follows. In the rest of this introduction, we discuss multiple connections between topics raised in this paper and physics. In Sect.~\ref{sect:DT_BSM}, we present a discrete-time, continuous-space version of the BSM model. An approach based on the formalism of Markov Decision Processes (MDP) is developed in 
Sect.~\ref{sect:MDP_BSM} where we introduce the QLBS model and obtain a DP-based solution for pricing and hedging in this model, that applies when the model is {\it known}.
A RL setting is discussed in Sect.~\ref{sect:QL_BSM}, where we introduce a finite-horizon Q-Learning and Fitted Q-Iteration (FQI) for our model, and derive RL-based pricing and hedging formulae that apply when the QLBS model goes {\it model-free} and purely data driven.   
In Sect.~\ref{sect:QL_BSM_numeric}, we briefly discuss possible numerical examples.
Finally, we conclude in Sect.~\ref{sect:Summary}.

\subsection{Links with physics}

The first successful probabilistic model of stock prices, the Arithmetic Brownian Motion (ABM) model of Bachelier 1900 \cite{Bachelier}, is mathematically identical to the theory of free diffusion that was developed by Einstein five years later in 1905. The BSM model uses the Geometric Brownian Motion (GBM) model of 
Samuelson \cite{Samuelson} that applies the ABM model to logarithms of stock prices rather than stock prices themselves. As the GBM model is known to produce only a very rough approximation to dynamics of real stock prices, this translates into inaccuracies of option prices computed within the BSM model. 
The model presented in this paper develops a framework that relies directly on the pricing and trading data, thus bypassing the need for a parametric diffusion model for the stock price. 

The model of this paper also makes another connection between option pricing and physics more transparent. 
Option prices obtained with the classical BSM model do not correspond to any minimization or maximization problem. 
The classical BSM model is a model of an 'equilibrium' option pricing that does not explain how this equilibrium is obtained (or why such redundant instruments as options would be needed). 
This is quite different from physics that is based on variational principles and amounts to various minimization problems. As will be shown below, the loss of an optimization part in the problem of option pricing happens do to the continuous time limit $ \Delta t \rightarrow 0 $ that is routinely made in the BSM model (and other models of mathematical finance) right {\it from the start}. This limit is shown to be degenerate for the option pricing problem, because the objective function ceases to exist in this limit. On the other hand, for a discrete-time case $ \Delta t > 0 $ which corresponds to ``physical" option markets where the time step $ \Delta t $ is chosen according to the re-hedging frequency, the option pricing is argued to amount to an optimization problem that is similar to variational methods of Hamiltonian mechanics and stochastic HJB optimal control in continuous time.   


\section{Discrete-time Black-Scholes-Merton model}
\label{sect:DT_BSM}

We start with a discrete-time version of the BSM model. As is well known, the problem of option hedging and pricing in this formulation amounts to a sequential risk minimization.
The main open question is {\it how} to define risk in an option. In this part, we follow a local risk minimization approach pioneered in the work of F{\"o}llmer and Schweizer \cite{FS}, Schweizer \cite{Schweizer}, and Cern{\'y} and Kallsen \cite{CK}. A similar method was developed by physicists  Potters and Bouchaud \cite{PB}, see also the work by Kapoor et. al. \cite{Kapoor}. We use a version of this approach suggested in a Ph.D. thesis by Grau \cite{Grau}.   

In this approach, we take the view of a seller of a European option (e.g. a put option) with maturity $ T $ and the terminal payoff of $ H_T(S_T) $ at maturity, that depends on a final stock price $ S_T $ at that time. To hedge the option, the seller use proceeds of the sale to set up a replicating (hedge) portfolio $ \Pi_t $ made of the stock $ S_t $ and a risk-free bank deposit $ B_t $. The value of hedge portfolio at any time $ t \leq T $ is 
\beq
\label{Pi_t}
\Pi_t =  u_t  S_t + B_t
\eeq  
where $ u_t  $ is a position in the stock at time $ t $, taken to hedge risk in the option. 

\subsection{Hedge portfolio evaluation}

As usual, the replicating portfolio tries to exactly match the option price in all possible future states of the world.  If we start at maturity $ T $ when the option position is closed, the hedge $ u_t  $ should be closed at the same time, i.e. all stocks should be converted into cash. If we neglect transaction costs, as we do in this paper, we can enforce it by simply setting $ u_T = 0 $, because the portfolio is self-financing (see below), and thus setting $ u_T = 0 $ will automatically convert all remaining stock into cash without any cost\footnote{For more details on the choice $ u_T = 0 $, see \cite{Grau}.}.  Therefore, we set
\beq
\label{B_T}
\Pi_T = B_T =  H_T(S_T)
\eeq
which defines a terminal condition for $ B_T $ that should hold in all future states of the world at time $ T $.

To find an amount needed to be held in the bank account at previous times $ t < T $, we impose the self-financing constraint which requires that all future changes in the hedge portfolio should be funded from an initially set bank account, without any cash infusions or withdrawals over the lifetime of the option. This implies the following relation that ensures conservation of the portfolio value by a re-hedge at time $ t +1 $:
\beq
\label{self-financing}
 u_t  S_{t+1} + e^{r \Delta t} B_t = u_{t+1} S_{t+1} +  B_{t+1}
 \eeq
 This can be expressed as a recursive relation to calculate the amount of money to keep at the bank account to hedge the option at any time $ t < T $ using its value at the next time instance:
 \beq
 \label{B_t}
 B_t  = e^{- r \Delta t} \left[ B_{t+1} + \left( u_{t+1} - u_{t} \right) S_{t+1} \right] \, , \; \; \; t = T-1,\ldots, 0
 \eeq
Plugging this into Eq.(\ref{Pi_t}) produces a recursive relation for $ \Pi_t $ in terms of its values at later times, which can therefore be solved backward in time, starting from $ t = T $ with the terminal condition (\ref{B_T}), and continued all the way to the current time $ t = 0 $:
 \beq
 \label{Pi_t_rec}
 \Pi_t = e^{-r \Delta t} \left[ \Pi_{t+1} - u_t   \Delta S_t \right] , \; \; \;  \Delta S_t = S_{t+1} 
 - e^{ r \Delta t} S_t  \, , \; \; \; t = T-1,\ldots, 0 
 \eeq 
Note that Eqs.(\ref{B_t}) and (\ref{Pi_t_rec}) imply that both $ B_t $ and $ \Pi_t $ are not measurable at any $ t  < T $, as  they depend on the future. Respectively, their values today $ B_0 $ and $ \Pi_0 $ will be random quantities with some distributions. For any given hedging strategy $ \left\{ u_t  \right\}_{t=0}^{T} $, these distributions can be estimated using Monte Carlo simulation, which first simulates $ N $ paths of the underlying $ S_1 \rightarrow 
S_2 \rightarrow \ldots \rightarrow S_N $, and then evaluates $ \Pi_t $ going backward on each path. 
Note that because the choice of a hedge strategy does not affect the evolution of the underlying, such simulation of forward paths should only be performed once, and then re-used for future evaluations of the hedge portfolio under difference hedge strategy scenarios.
Alternatively, the distribution of the hedge portfolio value $ \Pi_0 $ can be estimated using real historical data for stock prices, together with a pre-determined hedging strategy   $ \left\{ u_t  \right\}_{t=0}^{T} $ and a terminal condition (\ref{B_T}). 

To summarize, the forward pass of Monte Carlo simulation is done by simulating the process $ S_1 \rightarrow 
S_2 \rightarrow \ldots \rightarrow S_N $,  while the backward pass is performed using the recursion (\ref{Pi_t_rec}) that takes a prescribed hedge strategy $ \left\{ u_t  \right\}_{t=0}^{T} $ and back-propagates uncertainty in the future into uncertainty today, via the self-financing constraint (\ref{self-financing}) \cite{Grau} which serves as a 'time machine for risk'.

This is exactly what the dealer (seller of the option) needs, as she has to set the price of the option {\it today}. This can be done e.g. by setting the option price to be the mean of the distribution of $ \Pi_0 $, plus some premium for risk.
But all this should obviously come only {\it after} the seller decides on a {\it hedging strategy}  $ \left\{ u_t  \right\}_{t=0}^{T} $ to be used in the future, that would be applied in the same way (as a mapping) for any future value   $ \left\{ \Pi_t \right\}_{t=0}^{T} $.
The choice of an {\it optimal} hedge strategy  $ \left\{ u_t  \right\}_{t=0}^{T} $ will therefore be discussed next.

\subsection{Optimal hedging strategy}
\label{sect:Optimal_hedging_strategy}

Unlike the recursive calculation of the hedge portfolio value (\ref{Pi_t_rec}) which is done {\it path-wise}, optimal hedges are computed using a {\it cross-sectional} analysis that operates simultaneously on all Monte Carlo paths. 
This is because we need to learn a {\it strategy} $ \left\{ u_t  \right\}_{t=0}^{T} $ that would apply to {\it all} states that might be 
encountered in the future, but each given MC path only produces {\it one} value $ S_t $ at time $ t $. Therefore, to compute an optimal hedge $ u_t(S_t) $ for a given time step $ t $, we need a {\it cross-sectional} information on {\it all} Monte Carlo paths at this time.  

Similar to the portfolio value calculation, the optimal hedges   $ \left\{ u_t  \right\}_{t=0}^{T} $ are computed backward in time, starting from $ t = T $. However, because we cannot know the future when we compute a hedge, for each time $ t $, any calculation of an optimal hedge $ u_t  $ can only condition on the information $ \mathcal{F}_t $ available at time $ t $. This is similar to the American Monte Carlo method of Longstaff and Schwartz \cite{LS} (see Appendix B for a short overview).

The optimal hedge $ u^{\star}(S_t) $ in this model is obtained from the requirement that the variance of $ \Pi_t $ across all simulated MC paths at time $ t $ is minimized when conditioned on the currently available {\it cross-sectional} information $   \mathcal{F}_t $, i.e. 
\bea
\label{opt_theta}
 u_t^{\star}(S_t) &=& arg \min_{u} Var \left[ \Pi_t | \mathcal{F}_t \right]   \nonumber \\
 &=& arg \min_{u} Var \left[ \Pi_{t+1} - u_t  \Delta S_t  | \mathcal{F}_t \right]  \, , \; \; \; t = T-1,\ldots, 0
 \eea
 Note the the first expression in Eq.(\ref{opt_theta}) implies that all uncertainty in $ \Pi_t $ is due to uncertainty regarding the amount $ B_t $ needed to be held at the bank account at time $ t $ in order to be able to cover the future obligations at the option maturity $ T $. This means that an optimal hedge should minimize the cost of a hedge capital for the option position at each time step $ t $. 
 
 The optimal hedge can be found analytically by setting the derivative of (\ref{opt_theta}) to zero. This gives
 \beq
 \label{phi_star}
  u_t^{\star}(S_t) = \frac{ Cov \left( \left. \Pi_{t+1}, \Delta S_t  \right| \mathcal{F}_t \right)}{
  Var \left( \left. \Delta S_t  \right| \mathcal{F}_t \right)} \, , \; \; \; t = T-1,\ldots, 0
  \eeq
This expression involves one-step expectations of quantities at time $ t + 1 $, conditional on time $ t $. 
How they can be computed depends on whether we deal with a continuous or a discrete state space.
If the state space is discrete, then such one-step conditional expectations are simply finite sums involving transition probabilities 
of a MDP model. If, on the other hand, we work in a continuous-state setting, these conditional expectations can be calculated 
in a Monte Carlo setting by using expansions in basis functions, similarly to the LSMC method of Longstaff and Schwartz 
\cite{LS}, or real-measure MC methods of Refs.~\cite{PB, Grau, Kapoor}.

In our exposition below, we use a general notation as in Eq.(\ref{phi_star}) to denote similar conditional expectations
where $ \mathcal{F}_t $ stands for cross-sectional information set at time $ t $, which lets us keep the formalism general enough to handle both cases of a continuous and a discrete state spaces, and discuss simplifications that arise in a special case of 
a discrete-state formulation separately, whenever appropriate. 


\subsection{Option pricing in discrete time}
\label{sect:Opt_pricing_discrete_time}

We start with the notion of a {\it fair} option price $ \hat{C}_t $ defined as a time-$t$ expected value of the hedge portfolio $ \Pi_t $:
\beq
\label{hat_C_t}
\hat{C}_t =  \mathbb{E}_t \left[  \left. \Pi_t  \right| \mathcal{F}_t \right]
\eeq
Using Eq.(\ref{Pi_t_rec}) and the tower law of conditional expectations, we obtain
\bea
\label{C_t_rec}
\hat{C}_t &=& \mathbb{E}_t \left[ \left. e^{-r \Delta t} \Pi_{t+1} \right| \mathcal{F}_t \right]
- u_t(S_t)  \, \mathbb{E}_t \left[ \left.  \Delta  S_{t}  \right| \mathcal{F}_t \right]  \nonumber \\
&=&  \mathbb{E}_t \left[ \left. e^{-r \Delta t}  \mathbb{E}_{t+1} \left[ \left. \Pi_{t+1} \right| \mathcal{F}_{t+1} \right] 
\right| \mathcal{F}_t \right] 
- u_t(S_t)  \, \mathbb{E}_t \left[ \left.  \Delta  S_{t}  \right| \mathcal{F}_t \right]  \\
&=& \mathbb{E}_t \left[ \left. e^{-r \Delta t} \hat{C}_{t+1} \right| \mathcal{F}_t \right] 
- u_t(S_t)  \, \mathbb{E}_t \left[ \left.  \Delta  S_{t}  \right| \mathcal{F}_t \right]  \, , \; \; \; t = T-1,\ldots, 0 \nonumber
\eea
Note that we can similarly use the tower law of conditional expectations to express the optimal hedge in terms of $ \hat{C}_{t+1} $ instead of $ \Pi_{t+1} $: 
 \beq
 \label{phi_star_2}
  u_t^{\star}(S_t) = \frac{ Cov \left( \left. \Pi_{t+1}, \Delta S_t  \right| \mathcal{F}_t \right)}{
  Var \left( \left. \Delta S_t  \right| \mathcal{F}_t \right)} 
  = \frac{ Cov \left( \left. \hat{C}_{t+1}, \Delta S_t  \right| \mathcal{F}_t \right)}{
  Var \left( \left. \Delta S_t  \right| \mathcal{F}_t \right)}
  \eeq
If we now substitute (\ref{phi_star_2}) into (\ref{C_t_rec}) and re-arrange terms, we can put the recursive relation for 
$ \hat{C}_t $ in the following form:
\beq
\label{signed_measure}
\hat{C}_t = e^{-r \Delta t} \mathbb{E}^{\hat{\mathbb{Q}}} \left[ \left. \hat{C}_{t+1} \right| \mathcal{F}_t \right]\, , \; \; \; t = T-1,\ldots, 0
\eeq
where $ \hat{\mathbb{Q}} $ is a signed measure with transition probabilities
\beq
\label{signed_measure_probs}
\tilde{q} \left( \left. S_{t+1}  \right| S_t \right) = p \left( \left. S_{t+1}  \right| S_t \right)  \left[ 1
- \frac{ \left( \Delta S_t - \mathbb{E}_t \left[ \Delta S_t \right] \right)  \mathbb{E}_t \left[ \Delta S_t \right]}{
Var \left( \left.  \Delta S_t \right| \mathcal{F}_t \right)} \right]
\eeq
where $  p \left( \left. S_{t+1}  \right| S_t \right) $ are transition probabilities under the physical measure $ \mathbb{P} $. 
Note that for large enough moves $ \Delta S_t $ this expression may become negative, which means that $ \hat{ \mathbb{Q}} $ is only a signed measure rather than a genuine probability measure. 

A potential for a negative fair option price $ \hat{C}_t $ is a well-known property of quadratic risk minimization schemes \cite{FS, Schweizer, CK, PB, Grau}. We do {\it not} view this as a problem for our model, because
the  ``fair" (expected) option price (\ref{hat_C_t}) is {\it not} a price a seller of the option should charge. 
The actual fair {\it risk-adjusted} price is given by Eq.(\ref{ask_price}) below, which can always be made non-negative by a proper level of risk aversion $ \lambda $ which is defined by the seller's risk preferences. In any case, our model does {\it not} rely on the no-arbitrage principle because it works directly with raw data. If data respect the no-arbitrage principle, our model will respect it too, by construction.  
 
Still, for the sake of completeness, an alternative approach that avoids negative option prices for arbitrary market/hedging scenarios is presented in Appendix A. Instead of a quadratic risk minimization, this approach relies on an  
exponential utility $ U(X) = - \exp( - \gamma X ) $ . 
As shown in Appendix A, the hedges and prices corresponding to the quadratic risk minimization scheme can be obtained with the exponential utility in 
the  limit of a small risk aversion $ \gamma \rightarrow 0 $, alongside calculable corrections via an expansion in powers of 
$ \gamma $.   
As our focus here is on a RL agent that  learns from {\it data} generated from an option trading strategy, 
in what follows we stick to a simple quadratic hedge (\ref{phi_star}).
If so desired, the whole scheme for optimal hedging and pricing can instead be constructed as described in Appendix A. 

Coming back to pricing of the option, the dealer cannot just ask the fair price $ \hat{C}_0 $ when selling the option, as she has to compensate for risk of exhausting the bank account $ B_t $ some time in the future, which would require cash infusions into the hedge portfolio, after any fixed amount $ \hat{B}_0 = \mathbb{E}_0 \left[ B_0 \right] $ is put in the bank account at time $ t = 0$ right after selling the option. One possible specification of a risk premium that the dealer has to add on top of the fair option price to come up with her own optimal ask price is to add the cumulative expected discounted variance of the hedge portfolio along all time steps $ t = 0, \ldots, N$, with a risk-aversion parameter $ \lambda $:
\beq
\label{ask_price}
 C_{0}^{(ask)} (S,u)  
= \mathbb{E}_0 \left[ \left. \Pi_0 + \lambda \sum_{t=0}^{T}   
e^{-r t} Var  \left[  \left. \Pi_t \right| \mathcal{F}_t \right]  \right| S_0 = S, u_0 = u \right] 
\eeq
Note that while the idea of adding an option price premium proportional to the variance of the hedge portfolio as done in Eq.(\ref{ask_price}) was initially suggested on the intuitive grounds by Potters and Bouchaud \cite{PB}, a utility-based approach presented in Appendix A actually {\it derives} it as a quadratic approximation to a utility-based option price, which also establishes 
an approximate relation between a risk aversion parameter $ \lambda $ 
of the quadratic risk optimization, and  a parameter $ \gamma $ of the exponential utility  $ U(X) = - \exp( - \gamma X ) $:
\beq
\label{lambda_gamma}
\lambda \simeq \frac{1}{2} \gamma 
\eeq 
For what follows, we note that the problem of {\it minimization} of a fair (to the dealer) option price (\ref{ask_price}) can be equivalently expresses as the problem of {\it maximization} of its negative $ V_t = - C_t^{(ask)} $, where
\beq
\label{Value_Function_port}
V_t( S_t) =   \mathbb{E}_t \left[    \left. - \Pi_{t}  
-   \lambda \, \sum_{t'=t}^{T}  e^{-r (t'-t)} Var \left[  \left. \Pi_{t'}   \right| \mathcal{F}_{t'} \right]  
\right|  \mathcal{F}_t \right]  
\eeq

\subsection{Hedging and pricing in the BS limit}
\label{sect:Hedging_BS_limit}

The framework presented above provides a smooth transition to the strict BS limit $ \Delta t \rightarrow 0 $. In this limit, the BSM model dynamics under the physical measure $ \mathbb{P} $ is described by a continuous-time Geometric Brownian motion with a drift $ \mu $ and volatility $ \sigma $:
\beq
\label{GBM}
\frac{d S_t}{S_t} = \mu dt + \sigma dW_t
\eeq
where $ W_t $ is a standard Brownian motion. 

Consider first the optimal hedge strategy (\ref{phi_star_2}) in the BS limit $ \Delta t \rightarrow 0 $. Using the first-order Taylor expansion
\beq
\label{FOTE}
  \hat{C}_{t+1} = C_t  +  \frac{\partial C_t}{\partial S_t} \Delta S_t + O \left( \Delta t \right) 
 \eeq
in (\ref{phi_star_2}), we obtain 
 \beq
 \label{theta_BS}
 u_t^{BS} \left( S_t \right) = \lim_{\Delta t \rightarrow 0}  u_t^{\star}(S_t)  =   \frac{\partial C_t}{\partial S_t} 
 \eeq
 which is the correct optimal hedge in the continuous-time BSM model.

To find the continuous-time limit of the option price,  we first compute the limit of the second term in Eq.(\ref{C_t_rec}):
\beq
\label{second_term}
\lim_{\Delta t \rightarrow 0} u_t(S_t)  \, \mathbb{E}_t \left[ \left.  \Delta  S_{t}  \right| \mathcal{F}_t \right]  = 
\lim_{ dt \rightarrow 0} u_t^{BS} S_t ( \mu - r) dt =  \lim_{dt \rightarrow 0} ( \mu - r) S_t \frac{\partial C_t}{\partial S_t} dt 
\eeq
To evaluate the first term in Eq.(\ref{C_t_rec}),  we use the second-order Taylor expansion:
\bea
\label{SOTE}
& & \hat{C}_{t+1} = C_t + \frac{\partial C_t}{\partial t} dt + \frac{\partial C_t}{\partial S_t} d S_t +  
\frac{1}{2}  \frac{\partial^2 C_t}{\partial S_t^2}  \left( d S_t \right)^2 + \ldots  \\
&& = C_t + \frac{\partial C_t}{\partial t} dt + \frac{\partial C_t}{\partial S_t}  S_t  \left( \mu dt + \sigma dW_t \right) +  
\frac{1}{2}  \frac{\partial^2 C_t}{\partial S_t^2}  S_t^2 \left( \sigma^2 dW_t^2 + 2  \mu \sigma dW_t  dt  \right) + O \left(dt^2 \right)
\nonumber 
\eea
Plugging Eqs.(\ref{second_term}) and (\ref{SOTE}) into Eq.(\ref{C_t_rec}), using $ \mathbb{E} \left[ d W_t \right] = 0 $ and 
$ \mathbb{E} \left[ d W_t^2 \right] = dt $, and simplifying, we find that  the stock drift $ \mu $ under the physical measure $ \mathbb{P} $ drops out from the problem, and Eq.(\ref{C_t_rec}) becomes the celebrated Black-Scholes equation in the limit $ dt \rightarrow 0 $:
\beq
\label{BS_eq}
 \frac{\partial C_t}{\partial t} + r S_t \frac{\partial C_t}{\partial S_t} + \frac{1}{2}  \sigma^2  S_t^2 \frac{\partial^2 C_t}{\partial S_t^2} 
 - r C_t = 0
 \eeq
 Therefore, if the world is lognormal, both our hedging and pricing formulae become the original formulae of the Black-Scholes-Merton model in the strict limit $ \Delta t \rightarrow 0 $.  
   
\section{QLBS }
\label{sect:MDP_BSM}

Now we will re-formulate the hedging and pricing method presented in Sect.~\ref{sect:DT_BSM} 
using the language of Markov Decision Processes (MDP). By keeping in line with formulae presented in Sect.~\ref{sect:DT_BSM}
that are valid for both a discrete-state and continuous-state cases, the MDP formulation to be presented here applies 
for both discrete-state and continuous-state spaces. 
  
Our discrete-time MDP model provides a discrete-time approximation to the hedging and pricing strategy of the BSM model, but by working directly in the physical measure $ \mathbb{P} $, and viewing the problem of option pricing and hedging as a problem of stochastic optimal control in discrete time, where the system being controlled is a hedge portfolio, and control is a stock position in this hedge portfolio. The problem is then solved by a sequential maximization of ``rewards". These ``rewards" are negatives of hedge portfolio one-step variances times the risk-aversion $ \lambda $, plus a drift term.  

As will be shown in detail below, when transition probabilities and a reward function are {\it known}, 
a Bellman optimality equation for this model can be solved 
using a recursive backward Value Iteration method. It turns out that a policy optimization step in this method can be 
performed {\it analytically} within our model, as this step amounts to a quadratic optimization problem.
The whole calculation is therefore semi-analytical, and only involves matrix linear algebra (linear regression) for a numerical implementation. 

On the other hand, we might know only the general {\it structure} of a MDP model, but {\it not} its specifications such as transition probability and reward function. In this case, we should solve      
a backward recursion for the Bellman optimality equation relying only on {\it samples} of data. This is a setting 
of {\it Reinforcement Learning}.  

It turns out that a Bellman optimality equation for our MDP model without knowing model dynamics by relying only on {\it data} can be easily solved (also semi-analytically, due to a quadratic reward function!) by 
the celebrated {\it Q-Learning} method of Watkins \cite{Watkins_1989, Watkins}.  
As will be shown below, a Q-Learner is {\it guaranteed} (by construction) to converge to the correct optimal hedge for any time step size, given enough training data, and is guaranteed to converge to the classical BSM results for option price and hedge ratio,
if dynamics are lognormal, in
the continuous time limit $ \Delta t \rightarrow 0 $\footnote{
Incidentally, this limit is also a non-interesting one due to its degeneracy. The continuous-time approach with diffusive price dynamics that involves both the classical BSM model and its multiple extensions, is only a model of a mean option price. Both mis-hedging risk and risk preferences of option traders are entirely missing out in this model of a market. This might be problematic as a framework for modeling real markets, as a market described by such approach is in a sense a 'fake' market -   it should not exist in the first place, for the exact same reason of absence of any risk, and thus any {\it incentives to trade} in such market.}.

\subsection{State variables}

As stock price dynamics typically involve a deterministic drift part, we can consider a change of state variable such that new, time-transformed variables would be stationary, i.e. non-drifting. For a given stock price process $ S_t $, we can achieve this by defining a new variable $ X_t $ by the following relation:
\beq
\label{X_t}
X_t = - \left( \mu - \frac{\sigma^2}{2} \right) t + \log S_t 
\eeq
This implies that 
\beq
\label{X_t_dt}
d X_t = - \left( \mu - \frac{\sigma^2}{2} \right) dt + d \log S_t = \sigma dW_t 
\eeq
Therefore, $ X_t $ is a standard Brownian motion, scaled by volatility $ \sigma $. If we know the value of $ X_t $ in a given MC scenario, the corresponding value of $ S_t $ is given by the formula 
\beq
\label{S_t_from_X_t}
S_t = e^{ X_t + \left( \mu - \frac{\sigma^2}{2} \right) t }
\eeq 
Note that as long as $ X_t $ is a martingale, i.e. $ \mathbb{E} \left[ d X_t \right] = 0 $, on average it should not run too far  away 
from an initial value $ X_0 $ during the lifetime of an option. The state variable $ X_t $ is time-uniform, unlike the stock 
price $ S_t $ that has a drift. But the relation (\ref{S_t_from_X_t}) can always be used in order to map non-stationary dynamics of 
$ S_t $ into stationary dynamics of $ X_t $. The martingale property of $ X_t $ is also helpful for numerical lattice approximations, as it implies that a lattice should not be too large to capture possible future variations of the stock price. 

The change of variables (\ref{X_t}) and its reverse (\ref{S_t_from_X_t}) can also be applied when the stock price dynamics are not GBM. Of course, the new state variable $ X_t $ will not in general be a martingale in this case, however is it intrinsically useful for
separating non-stationarity of the optimization task from non-stationarity of state variables. 

\subsection{Finite-state approximation}

Though most of our formalism to be presented below applies in both a continuous-state and discrete-state MDP settings, for 
benchmarking of discrete-state RL algorithms that are simpler than continuous-state algorithms, we may want to 
apply our method for a discrete-state setting at later stages.  This can be done by discretizing the set of admissible values of 
state variables $ X_t $ defined in Eq.(\ref{X_t}), while keeping the relation (\ref{S_t_from_X_t}) that expresses the stock
price $ S_t $ in terms of the (now discretized) state variable $ X_t $.  
For a  simple method to approximate a continuous-state BSM dynamics by a discrete-time, discrete-state Markov Chain model, see  Duan and Simonato \cite{DS}.
 
 As was mentioned in the introduction,  a state-space discretization is {\it not} necessary within the general 
methodology that we propose in this paper. A formalism presented below assumes a general continuous-state case in a Monte 
Carlo setting. This provides a very convenient way to research or benchmark {\it continuous-state} RL methods for financial applications. On the other hand, if we want to test {\it discrete-state} RL methods, a Markov Chain approximation to the BSM model can serve as a good benchmark model. 

Note that as long as the objective is to benchmark discrete-state RL algorithms, we do  
{\it not} necessarily have to worry about a numerical match between a discrete-time Markov Chain approximation and the original continuous-space BSM model. We can always simulate data {\it directly} from a Markov Chain model, with optimal hedges obtained in {\it this} model, and benchmark RL algorithms directly against {\it this model}, rather than against the original continuous-time BSM model.
 
Also note that in addition to providing a MDP-enabled formulation of the problem, a discrete-state approximation enables an easy computation of various one-step expectations entering our hedging and pricing formulas (\ref{phi_star}) and  (\ref{C_t_rec}).
Indeed, in this case these expectations would be given by finite sums over all feature discrete states reachable in one step from a given discrete state. In a continuous-space formulation with a MC setting as in Refs.~\cite{PB, Kapoor, Grau}, one has instead to rely on function approximations, using e.g. expansions in basis functions, similar to the American Monte Carlo method of Longstaff and Schwartz \cite{LS}, see also below.

\subsection{Value function and Bellman equations}

We start with re-stating the risk minimization procedure outlined above in Sect.~\ref{sect:Optimal_hedging_strategy} in a language of MDP problems. In particular, time-dependent state variables $ S_t $ are 
expressed in terms of time-homogeneous variables $ X_t $ using Eq.(\ref{S_t_from_X_t}). In addition, we 
will use the notation $ a_t = a_t (X_t) $ to denote actions expressed as functions of time-homogeneous 
variables $ X_t $. Actions $ u_t = u_t(S_t) $ in terms of stock prices are then obtained by the substitution
\beq
\label{action_map}
u_t \left( S_t \right) = a_t \left( X_t \left( S_t \right) \right) = a_t \left( \log S_t   - \left( \mu - \frac{\sigma^2}{2} \right) t \right) 
\eeq  
where we used Eq.(\ref{X_t}).

To differentiate between the actual hedging decisions $ a_t(x_t) $ where $ x_t $ is a particular realization of a random state 
$ X_t $ at time $ t $, and a hedging {\it strategy} that applies for any state $ X_t $, we introduce the notion of a time-dependent {\it policy} $ \pi \left(t, X_t \right) $. We consider deterministic policies, i.e.
\beq
\label{policy}
\pi: \left\{ 0, \ldots, T-1 \right\} \times \mathcal{X} \rightarrow \mathcal{A}
\eeq
is a deterministic policy that maps the time $ t $ and the current state $ X_t = x_t $ into the action $ a_t \in \mathcal{A} $:
 \beq
 \label{deterministic_policy}
 a_t = \pi(t, x_t) 
\eeq 
We start with the value maximization problem of Eq.(\ref{Value_Function_port}), which we re-write here in terms of a new state variable $ X_t $, and with an upper index to denote its dependence on the policy $ \pi $:  
\bea
\label{Value_Function_port_2}
& V_t^{\pi} ( X_t)  =   \mathbb{E}_t \left[    \left. - \Pi_{t}(X_t)  
-   \lambda \, \sum_{t'=t}^{T}  e^{-r (t'-t)} Var \left[  \left. \Pi_{t'}(X_{t'} )   \right| \mathcal{F}_{t'} \right]  
\right|  \mathcal{F}_t \right]  \nonumber \\
& =  \mathbb{E}_t \left[    \left. - \Pi_{t}(X_t)  - \lambda \, Var \left[ \Pi_t \right] 
-   \lambda \, \sum_{t'=t+1}^{T}  e^{-r (t'-t)} Var \left[  \left. \Pi_{t'}(X_{t'} )   \right| \mathcal{F}_{t'} \right]  
\right|  \mathcal{F}_t \right] 
\eea
The last term in this expression that involves a sum from $ t' = t+1 $ to  $ t' = T $ can be expressed in terms of $ V_{t+1} $ using the definition of the value function with a shifted time argument:
\beq
\label{last term}
- \lambda \mathbb{E}_{t+1} \left[ \sum_{t'=t+1}^{T} e^{-r(t' -t )} Var \left[ \left. \Pi_{t'} \right| \mathcal{F}_{t'} \right] \right]
= \gamma \left( V_{t+1} + \mathbb{E}_{t+1} \left[ \Pi_{t+1} \right] \right) \, ,  \; \; \gamma \equiv e^{-r \Delta t}
\eeq
Note that parameter $ \gamma $ introduced in the last relation is a discrete-time discount factor which in our framework is fixed in terms of a continuous-time risk-free interest rate $ r $ of the original BSM model. 

Substituting this into (\ref{Value_Function_port_2}), re-arranging terms and using the portfolio process Eq.(\ref{Pi_t_rec}), 
we obtain the Bellman equation for the QLBS model:
\beq
\label{MDP_BSM}
V_t^{\pi}(X_t) = \mathbb{E}_{t}^{\pi} \left[ R(X_t, a_t, X_{t+1}) + \gamma V_{t+1}^{\pi} \left( X_{t+1} \right) \right]
\eeq
where the one-step time-dependent random reward is defined as follows\footnote{Note that with our definition of the value function 
Eq.(\ref{Value_Function_port_2}), it is {\it not} equal to a discounted sum of future rewards.}:
\bea
\label{one-step-reward}
  R_t(X_t, a_t, X_{t+1})  &=&    \gamma a_t  \Delta S_t \left(X_t, X_{t+1} \right) - \lambda \, 
Var \left[ \left. \Pi_t  \right| \mathcal{F}_t  \right]  \, , \; \; t = 0,\ldots, T-1  \nonumber  \\
 & = & \gamma a_t \Delta S_t \left(X_t, X_{t+1} \right)   -  \lambda \gamma^2  
\mathbb{E}_t \left[  \hat{\Pi}_{t+1}^2  - 2 a_t \Delta {\hat S}_t \hat{\Pi}_{t+1} + a_t^2 \left( \Delta \hat{S}_t \right)^2  \right]  
\eea
where we used Eq.(\ref{Pi_t_rec}) in the second line, and $ \hat{\Pi}_{t+1} \equiv \Pi_{t+1} - \bar{\Pi}_{t+1} $, 
where $ \bar{\Pi}_{t+1} $ is the sample mean of all values of  $ \Pi_{t+1} $, and similarly for $  \Delta {\hat S}_t $.
For $ t = T $, we have $ R_T =  - \lambda \, Var \left[ \Pi_T \right]  $ where $ \Pi_T $ is determined by the terminal condition 
(\ref{B_T}).

Note that Eq.(\ref{one-step-reward}) implies that the expected reward $ R_t $ at time step $ t $  
is {\it quadratic} in the action variable  $ a_t $:
\beq
\label{quadratic_reward}
 \mathbb{E}_{t} \left[ R_t \left( X_t, a_t, X_{t+1} \right)  \right] =  \gamma a_t \mathbb{E}_t \left[ \Delta S_t \right] - \lambda \gamma^2  \mathbb{E}_{t} \left[ {\hat{\Pi}}_{t+1}^{2} - 2 a_t \Delta \hat{S}_t  \hat{\Pi}_{t+1}  + a_t^2 \left( \Delta \hat{S}_t \right)^2 \right]
\eeq
As we will see shortly, this proves very useful for a solution of the MDP dynamics in this model.
Also note that when $ \lambda \rightarrow 0 $, the expected reward is {\it linear} in $ a_t $, so it does not have a maximum.

As the one-step reward in our formulation incorporates variance of the hedge portfolio as a risk penalty, this approach belongs in a class of risk-sensitive
reinforcement learning. With our method, risk is incorporated to a traditional risk-neutral RL framework (which only aims at maximization of expected rewards) by 
modifying the one-step reward function.   
A similar construction of a risk-sensitive MDP by adding one-step 
variance penalties to a finite-horizon risk-neutral MDP problem was suggested in a different context by 
Gosavi \cite{Gosavi}.
 
The action-value function, or Q-function, is defined by an expectation of the same expression as in Eq.(\ref{Value_Function_port_2}), but 
 conditioned on both the current state $ X_t $ {\it and} the initial action $ a = a_t $, while following a policy $ \pi $ afterwards:  
\bea
\label{Value_maximization}
& Q_t^{\pi} (x,a) = 
  \mathbb{E}_t \left[ \left. - \Pi_{t}(X_t) \right| X_t = x, a_t = a \right]     \\
&-   \lambda \,  \mathbb{E}_t^{\pi} \left[  \left. \sum_{t'=t}^{T}  e^{-r (t'-t)} Var \left[  \left. \Pi_{t'}(X_{t'} )   \right| \mathcal{F}_{t'} \right]  
\right|  X_t = x, a_t = a \right]. \nonumber 
\eea

     
The optimal policy $ \pi_t^{\star} (\cdot | X_t) $ is defined as the policy which maximizes the 
value function $ V_t^{\pi} \left( X_t \right) $, or alternatively and equivalently, maximizes the action-value function $ Q_t^{\pi} \left( X_t, a_t \right) $:



\beq
\label{Value_maximization_pi}
\pi_t^{\star}(X_t) = \text{argmax}_{ \pi}  \, V_t^{\pi} ( X_t) = \text{argmax}_{a_t \in \mathcal{A}} Q_t^{\star} (X_t,a_t)
\eeq
The optimal value function satisfies the Bellman optimality equation
\beq
\label{Bellman_V_star}
V_t^{\star}(X_t) = \mathbb{E}_{t}^{\pi^{\star}} \left[ R_t(X_t, u_t = \pi_t^{\star}(X_t), X_{t+1}) + \gamma V_{t+1}^{\star} \left( X_{t+1} \right) \right].
\eeq
The Bellman optimality equation for the action-value function reads
\beq
\label{Bellman_Q}
Q_t^{\star} (x,a) =  \mathbb{E}_t \left[   R_t \left(X_t, a_t, X_{t+1} \right)   + \gamma \max_{a_{t+1} \in \mathcal{A}}  
 \left. Q_{t+1}^{\star} \left( X_{t+1}, a_{t+1} \right)  \right| X_t = x, a_t = a  \right]  \, , \; \; t = 0, \ldots, T-1,
\eeq
where a terminal condition at $ t = T $  
 \beq
 \label{Q_T}
 Q_T^{\star}(X_T, a_T = 0) = - \Pi_T \left(X_T \right) - \lambda \, Var \left[ \Pi_T \left( X_T \right) \right],
 \eeq 
 and where $ \Pi_T $ is determined by Eq.(\ref{B_T}). Recall that  $ Var \left[ \cdot \right] $  
 here means variance with respect to all Monte Carlo paths that terminate in a given state.

%


\subsection{Optimal policy}
\label{sect:Q_backward_recursion}


Substituting the expected reward (\ref{quadratic_reward}) into the Bellman optimality equation 
(\ref{Bellman_Q}), we obtain
\bea
\label{Bellman_Q_a_t}
  Q_{t}^{\star}(X_{t},a_{t}) &=& \gamma \mathbb{E}_{t} \left[ 
 Q_{t+1}^{\star} \left( X_{t+1}, a_{t+1}^{\star} \right) 
 + a_t \Delta S_t \right]   \nonumber \\
 &- &   \lambda \gamma^2   \, \mathbb{E}_t  \left[ \hat{\Pi}_{t+1}^{2}  - 2  a_t  \hat{\Pi}_{t+1} \Delta \hat{S}_t
 + a_t^2  \left( \Delta \hat{S}_t \right)^2  \right] \, , \; \; t = 0, \ldots, T-1.
\eea
Note that the first term $ \mathbb{E}_{t} \left[ 
 Q_{t+1}^{\star} \left( X_{t+1}, a_{t+1}^{\star} \right) \right]$ can depend on the current action only through the conditional probability $ p \left(X_{t+1} | X_{t} a_{t} \right) $. However, the next-state probability can depend on the current action, $ a_t $, only when there is a feedback loop of trading in option underlying stock on the stock price. 
 In the present framework, we follow the standard assumptions of the Black-Scholes model that assumes an option buyer or seller do not produce any market impact. 
 
Neglecting the feedback effect, the expectation $ \mathbb{E}_{t} \left[ Q_{t+1}^{\star} \left( X_{t+1}, a_{t+1}^{\star} \right) \right] $ does not depend on $ a_t $. Therefore, with this approximation, the action-value function $ Q_t^{\star} \left( X_t, a_t \right) $ is {\it quadratic} in the action variable  $ a_t $.

As  $ Q_t^{\star} \left( X_t, a_t \right) $  is a quadratic function of $ a_t $, the optimal action (i.e. the hedge) $ a_t^{\star} (S_t) $ that maximizes 
$ Q_{t}^{\star}(X_{t},a_{t}) $  is computed analytically:
\beq
\label{a_star_t}
a_{t}^{\star} \left( X_t \right)
 =  \frac{\mathbb{E}_{t} \left[  \Delta \hat{S}_{t}  \hat{\Pi}_{t+1} + \frac{1}{2 \gamma \lambda} \Delta S_{t} \right]}{
  \mathbb{E}_{t} \left[ \left( \Delta \hat{S}_{t} \right)^2 \right]} 
\eeq 
If we now take the limit of this expression as $ \Delta t \rightarrow 0 $  by using Taylor expansions around time $ t $ as 
in Sect.~\ref{sect:Hedging_BS_limit}, we obtain
\beq
\label{lim_a_star_t}
\lim_{\Delta t \rightarrow 0 } a_{t}^{\star} = \frac{ \partial \hat{C}_t }{\partial S_t} + \frac{ \mu - r}{ 2 \lambda \sigma^2} \frac{1}{S_t}
\eeq
Note that if we set $ \mu = r $, or alternatively if we take the limit $ \lambda \rightarrow \infty $, it becomes identical to the BS delta,
while the finite-$\Delta t $ delta in Eq.(\ref{a_star_t}) coincides in these cases with a local-risk minimization delta given by 
Eq.(\ref{phi_star}). Both these facts have related interpretations. 
The quadratic hedging that approximates option delta 
(see Sect.~\ref{sect:Hedging_BS_limit})  only looks at {\it risk} of a hedge portfolio, while here we extend it by adding a {\it 
drift term} $ \mathbb{E}_t \left[ \Pi_t \right] $ to
 the objective function, see Eq.(\ref{Value_Function_port}), in the spirit of Markowitz risk-adjusted portfolio return analysis \cite{Markowitz}. This produces a linear first term in the quadratic expected reward (\ref{quadratic_reward}).
 Resulting hedges are therefore different from hedges obtained by only minimizing risk.
Clearly, a pure risk-focused quadratic hedge corresponds to either taking the limit of {\it infinite} risk aversion rate in a Markowitz-like risk-return analysis, or setting $ \mu = r $ in the above formula, to achieve the same effect. Both factors appearing in 
Eq.(\ref{lim_a_star_t}) show these two possible ways to obtain pure risk-minimizing hedges from our more general hedges. 
Such hedges can be applied when an option is considered for investment/speculation, rather than only as a hedge instrument. 
 
 To summarize, the local risk-minimization hedge and fair price formulae of Sect.~\ref{sect:DT_BSM} are recovered from 
 Eqs.(\ref{a_star_t}) and (\ref{Bellman_Q_a_t}), respectively, if we first set $ \mu =  r $ in Eq.((\ref{a_star_t}), and then set 
 $ \lambda = 0 $ in Eq.(\ref{Bellman_Q_a_t}).  After that, the continuous-time BS formulae for these expressions are reproduced in the final limit $ \Delta t \rightarrow 0 $ in these resulting expressions, as discussed in Sec.~\ref{sect:DT_BSM}.
 Note that the order of taking the limits is to start with the hedge ratio (\ref{lim_a_star_t}), set there $ \mu = r $, then plug this into 
the price equation (\ref{Bellman_Q_a_t}), and take the limit $ \lambda \rightarrow 0 $ there. 
The latter relation yields the Black-Scholes equation in the limit $ \Delta t \rightarrow 0 $ as shown in 
Eq.(\ref{BS_eq}). This order of taking the BS limit is consistent with the principle of hedging first and pricing second, which is implemented in the QLBS model, as well as consistent  with market practices of working with illiquid options.  
   
Plugging Eq.(\ref{a_star_t}) back into Eq.(\ref{Bellman_Q_a_t}), we obtain an explicit 
recursive formula for the {\it optimal} action-value function:
\beq
\label{Q_star_rec}
Q_{t}^{\star}(X_{t},a_{t}^{\star}) = \gamma \mathbb{E}_{t} \left[ Q_{t+1}^{\star}(X_{t+1},a_{t+1}^{\star}) - \lambda \gamma 
\hat{\Pi}_{t+1}^2 + \lambda \gamma  \left( a_{t}^{\star} \left(X_t \right) \right)^2   \left( \Delta \hat{S}_{t} \right)^2  \right] 
 \, , \; \; t = 0, \ldots, T-1
\eeq
where $ a_{t}^{\star} \left( X_t \right) $ is defined in Eq.(\ref{a_star_t}). Note that this relation does {\it not} have the 
right risk-neutral limit when we set $ \lambda \rightarrow 0 $ in it. The reason is that  setting $ \lambda 
\rightarrow 0 $ in Eq.(\ref{Q_star_rec}) is equivalent to setting  $ \lambda 
\rightarrow 0 $ in Eq.(\ref{a_star_t}), but, as we just discussed, this would {\it not} be the right way to reproduce the BS option price equation (\ref{BS_eq}). 

The backward recursion given by Eqs. (\ref{a_star_t}) and  (\ref{Q_star_rec}) proceeds all the way backward starting at $ t = T - 1 $ to the present $ t = 0 $. At each time step, the problem of maximization over possible actions amounts to convex optimization 
which is done analytically using Eq.(\ref{a_star_t}), which is then substituted into Eq.(\ref{Q_star_rec}) for the current time step.
Note that such simplicity of action optimization in the Bellman optimality equation is not encountered very often in other SOC problems.
As Eq.(\ref{Q_star_rec}) provides the backward recursion directly for the {\it optimal} Q-function, neither continuous nor discrete action space representation is required in our setting, as the action in this equation is always just one {\it optimal} action.  
If we deal with a finite-state QLBS model, then the values of the optimal time-$t$ Q-function for each node are 
obtained directly from sums of values of the next-step expectation in various states at time $ t + 1 $, times one-step probabilities to reach these states.

The end result of the backward recursion for the action-value function is its current value. According to our definition of the option price (\ref{ask_price}), it is exactly the negative of the optimal $ Q $-function. We therefore obtain the following expression for the fair ask option price in our approach, which we can refer to as the {\it QLBS option price}: 
\beq
\label{ask_price_QLBS}
 C_{t}^{(QLBS)}(S_t, ask)  = - Q_t \left(S_t, a_t^{\star} \right) 
\eeq
It is interesting to note that while in the original BSM model the price and the hedge for an option are given by two separate expressions, in the QLBS model, they are parts of the {\it same} expression (\ref{ask_price_QLBS}) - simply 
because its option price is the (negative of the) 
optimal Q-function, whose second argument is by construction the optimal {\it action} - which corresponds to the optimal hedge in the setting 
of the QLBS model.
  
Eqs.(\ref{ask_price_QLBS}) and (\ref{a_star_t}) that give, respectively, the optimal price and the optimal hedge for the option, 
jointly provide a complete solution of the QLBS model (when the dynamics are {\it known}) that generalizes the classical BSM model towards a non-asymptotic case $ \Delta t > 0 $, while reducing to the latter in the strict BSM limit $ \Delta t \rightarrow 0 $. In the next section, we will see how they can be implemented.

\subsection{DP solution: Monte Carlo implementation}
\label{sect:DP}

In practice, the backward recursion expressed by Eqs.(\ref{a_star_t}) and (\ref{Q_star_rec}) is solved in a Monte Carlo setting, where we assume to have access to 
$ N_{MC} $ simulated (or real) paths for the state variable $ X_t $. 
In addition, we  assume that we have chosen a set of basis functions  $ \{ \Phi_n(x) \} $.

 Similar to Refs.~\cite{LS, PB, Grau}, our model uses {\it all} Monte Carlo (or historical) paths for the replicating portfolio simultaneously.
Note a difference between such approach and a lattice-based approach that could alternatively be used for a finite-state version of the model in the backward recursion for the Q-function described in 
Sect.~\ref{sect:Q_backward_recursion} . With a lattice approach, for each time step $ t $, expectations with respect to future scenarios at time $ t + 1 $ are computed for each node $ X_t $ separately. While such approach is fine when the model is known and a state space is finite and small, in a setting of RL, when the model is {\it unknown}, this amounts 
to asynchronous updates of both the policy and the action-value function, which might substantially slow down the learning process. 

However, in a Monte Carlo scheme, we average over all scenarios at time  $ t $ and $ t + 1 $ simultaneously, by taking an empirical mean of different path-wise Monte Carlo scenarios. Therefore there is no need in explicit conditioning on the market state $ X_t $ 
at time $ t $.  
State variable values at both times $ t $ and $ t + 1 $ 
for {\it all} MC paths are used {\it simultaneously} to find the optimal action and optimal Q-function at time $ t $, for all states 
$ X_t $. Learning optimal actions for all states simultaneously means learning a {\it policy}, which is exactly our objective. 

We can then expand  the optimal action (hedge)  $ a_t^{\star} \left( X_t \right)  $ and 
optimal Q-function $ Q_t^{\star} \left(X_t, a_t^{\star} \right) $ 
 in basis functions, with time-dependent coefficients:
\beq
\label{Q_basis_exp}
a_t^{\star} \left( X_t \right) = \sum_{n}^{M}  \phi_{nt} \Phi_n \left( X_t  \right) \, ,  \; \;  
Q_t^{\star} \left(X_t, a_t^{\star} \right) = \sum_{n}^{M} \omega_{nt} \Phi_n \left( X_t  \right)  
\eeq
Coefficients $ \phi_{nt} $ and $ \omega_{nt} $ are computed recursively backward in time for $ t = T-1, \ldots, 0 $.
First, we find coefficients $ \phi_{nt} $  of the optimal action expansion. This is found by minimization of the following 
quadratic functional that is obtained by replacing the expectation  in Eq.(\ref{Bellman_Q_a_t}) by a MC estimate,
dropping all $ a_t $-independent terms, substituting the expansion (\ref{Q_basis_exp}) for $ a_t $, 
and changing the overall sign to convert maximization into minimization:
\beq
\label{phi_n}
G_t(\phi) = \sum_{k=1}^{N_{MC}} \left(  
- \sum_{n}  \phi_{nt} \Phi_n \left( X_t^k  \right) \Delta S_t^k 
+ \gamma \lambda \left( \hat{\Pi}_ {t+1}^k - \sum_{n}  \phi_{nt} \Phi_n \left( X_t^k  \right) 
\Delta \hat{S}_t^k \right)^2 \right)  
 \eeq
 This formulation automatically takes care of averaging over market scenarios at time $ t $.
 
Minimization of Eq,(\ref{phi_n}) with respect to coefficients $ \phi_{nt} $ produces a set of linear equations:
\beq
\label{lin_eq_phi}
\sum_{m}^{M} A_{nm}^{(t)} \phi_{mt} = B_n^{(t)} \, , \; \; n = 1, \ldots, M
\eeq
where 
\bea
\label{AB}
A_{nm}^{(t)} &=& \sum_{k=1}^{N_{MC}} \Phi_n \left( X_t^k \right)  \Phi_m \left( X_t^k \right) \left(  \Delta \hat{S}_t^k  \right)^2 
\nonumber \\
B_{n}^{(t)} &=& \sum_{k=1}^{N_{MC}} \Phi_n \left( X_t^k \right)  \left[ \hat{\Pi}_ {t+1}^k \Delta \hat{S}_t^k
+ \frac{1}{2 \gamma \lambda} \Delta S_t^k  \right]
\eea
which produces the solution for the coefficients of expansions of the optimal action $ a_t^{\star} \left(X_t \right) $ in a vector form:
\beq
\label{phi_nt_vec}
{\bf \phi}_t^{\star} = {\bf A}_t^{-1} {\bf B}_t
\eeq 
where $ \bf{A}_t $ and $ \bf B_t $ are a matrix and vector, respectively, with matrix elements given by Eq.(\ref{AB}).
Note a similarity between this expression and the general relation (\ref{a_star_t}) for the optimal action.
%

Once the optimal action $ a_t^{\star} $ at time $ t $ is found in terms of its coefficients (\ref{phi_nt_vec}), we turn 
to the problem of finding coefficients $ \omega_{nt} $ of the basis function expansion  (\ref{Q_basis_exp}) for 
the optimal Q-function.  
To this end, the one-step Bellman optimality equation (\ref{Bellman_Q}) for  $ a_t = a_t^{\star} $
is interpreted as  regression of the form
\beq
\label{Q_star_rec_3}
 R_t \left(X_t, a_t^{\star}, X_{t+1} \right)   + \gamma \max_{a_{t+1} \in \mathcal{A}}  
  Q_{t+1}^{\star} \left( X_{t+1}, a_{t+1} \right)  =  Q_{t}^{\star}(X_{t},a_{t}^{\star}) + \varepsilon_t
\eeq
where $ \varepsilon_t $ is a random noise at time $ t $ with mean zero.
Clearly, taking expectations of both sides of (\ref{Q_star_rec_3}), we recover Eq.(\ref{Bellman_Q}) with $ a_t = a_t^{\star} $,
 therefore Eqs. (\ref{Q_star_rec_3}) and (\ref{Bellman_Q}) are equivalent in expectations when $ a_t = a_t^{\star} $. 

Coefficients $ \omega_{nt} $ are therefore found by solving the following least-square optimization problem:
\beq
\label{omega_problem}
F_t(\omega) = \sum_{k=1}^{N_{MC}} \left(  R_t \left(X_t, a_t^{\star}, X_{t+1} \right)   + \gamma \max_{a_{t+1} \in \mathcal{A}}  
  Q_{t+1}^{\star} \left( X_{t+1}, a_{t+1} \right)  - 
\sum_{n}^{M} \omega_{nt} \Phi_n 
\left( X_t^k \right) \right)^2 
\eeq 
Introducing another pair of a matrix $ \bf{C}_t $ and a vector $ \bf{D}_t $ with elements
\bea
\label{CD}
C_{nm}^{(t)} &=& \sum_{k=1}^{N_{MC}} \Phi_n \left( X_t^k \right)  \Phi_m \left( X_t^k \right)
\nonumber \\
D_{n}^{(t)} &=& \sum_{k=1}^{N_{MC}} \Phi_n \left( X_t^k \right)  
 \left(   R_t \left(X_t, a_t^{\star}, X_{t+1} \right)   + \gamma \max_{a_{t+1} \in \mathcal{A}}  
  Q_{t+1}^{\star} \left( X_{t+1}, a_{t+1} \right)   \right)
\eea
we obtain the vector-valued solution for optimal weights $ \omega_t $ defining the optimal Q-function at time $ t $:
\beq
\label{omega_nt_vec}
\omega_{t}^{\star} =  {\bf C}_t^{-1} {\bf D}_t
\eeq
Equations (\ref{phi_nt_vec}) and (\ref{omega_nt_vec}), computed jointly and recursively for $ t = T-1, \ldots, 0 $
provide a practical implementation of the backward recursion scheme of Sect.~\ref{sect:Q_backward_recursion} in a continuous-space setting using expansions in basis functions. This approach can be used to find optimal price and optimal hedge when the dynamics are {\it known}.

%

Note that equations (\ref{phi_nt_vec})-(\ref{omega_nt_vec}) can be used for both continuous-state and finite-state MDP setting. If one works with a 
discrete-state version of QLBS, we can simply use ``one-hot" basis functions, that are equal one for a given node, and zero otherwise:
\beq
\label{basis_lattice}
 \Phi_n^{(finite-state)} \left( X_t \right) = \delta_{ X_t, X_n}  
 \eeq
 where $ \delta_{a,b} = 1 $ if $ a = b $, and zero otherwise , and $ X_n $ is the $ n $-th node on a grid of discretized $ X $-values.  These basis functions would serve as accumulators of contributions of individual grid states into sums in Eqs.(\ref{AB}) and (\ref{CD}).
 A smoothed-out choice of such localized basis for a continuous-state version of our model could be provided by B-splines or Gaussian kernels, while for a multi-dimensional continuous-state case one could use multivariate B-splines or Radial Basis Functions (RBF's) \cite{QLBS_NuQLear}.

\subsection{Discussion}
\label{sect:Discussion}


 Our numerical scheme bears some similarities to the previous Monte Carlo approaches to option pricing.
Similar to the LSMC method of Longstaff and Schwartz
\cite{LS})  and the method of Grau \cite{Grau} (but different from the HMC method of Potters and Bouchaud \cite{PB}), 
with our approach, possible inaccuracies incurred in rolling back 
the optimal Q-function do {\it not} impact the resulting optimal hedge (\ref{phi_nt_vec}). This is similar to the LSMC method, where inaccuracies in determining the continuation value for an option do not impact the quality of a computed exercise boundary. 
However,  despite apparent similarities, the previous Monte Carlo based approaches do not offer consistent  
distribution-free methods for both option pricing {\it and} hedging. 

To illustrate this point, we note that if we set $ \mu = r $ in Eq.(\ref{a_star_t}), our hedge ratios become 
identical to those suggested by Grau \cite{Grau}, 
yet our model differs from his method in the {\it pricing} part. 
While Grau suggested to use a CVAR-adjusted price as a compensation for risk in the option \cite{Grau},
it is in a sense an ``after-thought" of his scheme. Another risk-related premium (for example, VAR) could be used there in place of CVAR, and the hedging method would still be the same. In other words, in this method, there is no direct link between hedging and pricing. Grau's approach is a pure
``direct policy search", in the language of Reinforcement Learning \cite{SB}, that does not rely on any value function in optimization of a hedge strategy. 

On the other hand, we suggest a different definition of a fair ask option price, that is entirely consistent with a prescribed local-risk minimization hedging strategy. This price is given by a negative of the Q-function 
$ Q_0(X_0, a_0^{\star}) $ at time $ t = 0 $. If the state dynamics and reward functions are {\it known}, then the Bellman optimality equation for the $ Q$-function can be solved recursively backward in time, starting from $ t = T-1 $. The optimization step, performed  $ T $ times in our scheme, amounts to a quadratic optimization task which can be performed analytically.
No discretization of the action space is required, due to a quadratic dependence of the action-value function $ Q_t  $ at time $ t $
on the action $ a_t $ for this time step. This implies, in particular, that we can benchmark both continuous-state and discrete-state RL algorithms with either a continuous-state (discrete-time) QLBS model, or a finite-state QLBS model which is obtained by a state-discretization of the continuous-state MDP model.   

%

The second comment we want to make is that our Bellman optimality equation (\ref{Bellman_Q}) for the optimal action-value function  
is {\it guaranteed} by construction, at least for a finite-state formulation of the QLBS model, to converge to the correct (negative of the) option price and option hedge of the BSM 
model in the joint limit  $  \lambda \rightarrow 0 , \, \Delta t \rightarrow 0  $, while providing calculable corrections for 
a pre-asymptotic regime $ \Delta t > 0, \,  \lambda > 0 $ by using a backward recursion of the $ Q$-function (\ref{Bellman_Q}).  
We will leave a detailed investigation of empirical behavior of option prices and hedges in this pre-asymptotic regime to a future work, while concentrating in this paper on a mathematical framework.

For a finite $ \Delta t $, finite $ \lambda $ regime, we have shown that a Bellman optimality equation for the $Q$-function
 provides a very natural and easily computable link between a particular local-risk minimization hedging strategy 
 (an optimal policy, in the language of MDP models) and a corresponding price for the option that should be a functional of this strategy. In our model, we suggested using a discounted expected cumulative variances of a hedge portfolio from all future hedge rebalance periods as a measure of expected risk in the option, that should be compensated to the dealer in accord with her risk tolerance $ \lambda $. While this choice of a risk-premium for option {\it pricing} is indeed non-unique, in  
 complete agreement with the principle 'hedging first, pricing next', as we saw above, it leads to a very tractable MDP formulation that can be solved semi-analytically using the backward recursion for the optimal $ Q$-function as described above, if both the dynamics and reward function are known.
 
 Convergence of the QLBS model to the BSM model implies that for a specific investment portfolio made of a stock and a bank cash account, our multi-period extension of the Markowitz theory converges to the BSM price and delta of the option. 
This link between the Markowitz portfolio model and the BSM model appears new, at least to the author.


\section{Q-Learning and Fitted Q Iteration in QLBS}
\label{sect:QL_BSM}

Reinforcement Learning (RL) solves the same problem as Dynamic Programming (DP), i.e. it finds an optimal policy.
But unlike DP, Reinforcement Learning does {\it not} assume that transition probabilities and reward function are known.
Instead, it relies on samples to find an optimal policy. We will focus on value-based RL methods that work with the same objects
(the value function and action-value function) as DP \cite{SB}. 

Our setting assumes a {\it batch-mode} learning, when we only have access to some historically collected data.
No access to a real-time environment, or a simulator of such environment, is assumed available. 
The data available is given by a set of $ N_{MC} $ trajectories for the underlying stock $ S_t $ (expressed as a function of $ X_t $ 
using Eq.(\ref{S_t_from_X_t})), hedge position $ a_t $ , instantaneous reward $ R_t $, and the next-time value $ X_{t+1} $:
\beq
\label{F_t_RL}
\mathcal{F}_t^{(n)} = \left\{  \left( X_t^{(n)}, a_t^{(n)}, R_t^{(n)}, X_{t+1}^{(n)} \right) \right\}_{t=0}^{T-1} \, ,  \; \; n = 1, \ldots, N_{MC} 
\eeq
Moreover, as long as the dynamics are Markov, a collection of $ N_{MC} $ trajectories of length $ T $ each is equivalent, for the purpose of training a RL model,  to a dataset of $ N_{MC} \times T $ single-step transitions. We assume that such dataset is available either 
as a simulated data, or as a real historical stock price data, combined with some artificial data that would track a performance of a hypothetical stock-and-cash replicating portfolio for a given option.
        
Neither the dynamics nor the true reward distribution are assumed known in  
such off-line setting for RL which is known as batch-mode Reinforcement Learning \cite{Ernst}. In particular, Ernst {\it et. al.} proposed to use a {\it fitted Q-iteration} method for such setting \cite{Ernst}. This method combines the celebrated Q-Learning method for model-free RL \cite{Watkins} with function approximations to treat large discrete or continuous state spaces. While Ernst  {\it et. al.} deal with infinite-horizon batch mode RL, this setting was extended to a finite-horizon case by Murphy \cite{Murphy}, see also a Ph.D. thesis by Fonteneau \cite{Fonteneau}.  We will start with a pure Q-Learning, and then discuss Fitted Q Iteration.

\subsection{Q-Learning}
\label{sect:Q_Learning}

Here we assume a  discrete-state/discrete-action version of the QLBS model, as the Q-Learning in its original form suggested by Watkins in 1989 \cite{Watkins_1989} applies only for such setting. In this case,  Q-Learning converges to the true optimal action-value 
function with probability one, given enough data. Note that the original Q-Learning method of Watkins \cite{Watkins_1989, Watkins} refers to an infinite-horizon MDP problem, while we deal with a {\it finite-horizon} MDP problem. But stochastic approximation works the same way in this setting as for infinite-horizon problems: we just have to apply it separately at each step of a backward recursion for the Bellman optimality equation (\ref{Bellman_Q}) for the $Q$-function, and asymptotic convergence is still guaranteed 
\cite{Murphy}.

The classical Q-Learning method implements a single-step update of a current value of the $Q$-function, or in fact, a 
$Q$-{\it value} for each combination of a state and action, as both the state space and action space are discrete. In other words, the $Q$-function for an infinite-horizon discrete-state/discrete-action MDP is represented in a {\it tabulated form}, as a two-dimensional matrix/tensor. For a {\it time-dependent} problem with a {\it finite} time horizon, which is our case here, a time argument is added to a $ Q $-function/table, so it is represented by a {\it three}-dimensional tensor for a discrete state-action space. 

Q-Learning is obtained by using the Robbins-Monro stochastic approximation (see e.g. \cite{Gosavi_2014}) to estimate the unknown expectation in  Eq.(\ref{Bellman_Q}) which we repeat here for convenience:
\beq
\label{Bellman_Q_RL}
Q_t^{\star} (x,a) =  \mathbb{E}_t \left[   R_t \left(X_t, a_t, X_{t+1} \right)   + \gamma \max_{a_{t+1} \in \mathcal{A}}  
 \left. Q_{t+1}^{\star} \left( X_{t+1}, a_{t+1} \right)  \right| X_t = x, a_t = a  \right]  \, , \; \; t = 0, \ldots, T-1
\eeq
with a terminal condition at $ t = T $ given by Eqs.(\ref{Q_T}) and (\ref{B_T}).
Recall that  $ Var \left[ \cdot \right] $  
here means variance with respect to all Monte Carlo paths that end up in a given state. 
 
Using the Robbins-Monro approximation for the expectation in (\ref{Bellman_Q_RL}), we obtain an update rule for the optimal Q-function with online Q-Learning
after observing one datapoint $  \left( X_t^{(n)}, a_t^{(n)}, R_t^{(n)}, X_{t+1}^{(n)} \right) $:
\beq
\label{Q_learning}
Q_{t}^{\star, k+1}(X_{t},a_{t}) = (1 - \alpha^k)  Q_{t}^{\star, k}(X_{t},a_{t}) + \alpha^k 
\left[ R_t \left(X_t, a_t, X_{t+1} \right) +  \gamma \max_{a_{t+1} \in \mathcal{A}}  
  Q_{t+1}^{\star,k} \left( X_{t+1}, a_{t+1} \right) \right]  
\eeq
In addition to an online update of the Q-function, similar Robbins-Monro updates should be applied to estimate expectations that enter Eq.(\ref{a_star_t}). 

The Q-iteration rule of Eq.(\ref{Q_learning}) shows why Q-Learning is both a {\it model-free} and {\it off-policy} algorithm: it 
{\it does not make any assumptions} on 
a true data-generating process that produced the observation $  \left( X_t^{(n)}, a_t^{(n)}, R_t^{(n)}, X_{t+1}^{(n)} \right) $.
It simply takes it as {\it given}, and updates the action-value function for a given state-action node $ \left( X_t^{(n)}, a_t^{(n)} \right) $.
Q-iteration (\ref{Q_learning}) is guaranteed to asymptotically converge to a true optimal value function for all state-action pairs, given that each such pair is encountered infinitely many times in data \cite{Watkins_1989}. 
For more on convergence of Q-Learning algorithms see e.g. \cite{Gosavi_2014}.

Even though the classical online $Q$-Learning algorithm is guaranteed to asymptotically converge, it might take it too long for practical purposes. The reason for this is that 
optimal hedges are obtained using {\it cross-sectional} information across all Monte Carlo paths. Such cross-sectional information would be masked for any on-line method of updating $ Q $-values and optimal hedge ratios $ a_t^{\star} \left(X_t \right) $ on a space grid. 

However, the way out in our case is also clear - we simply have to update $ Q $-values and values of $ a_t^{\star} \left(X_t \right) $ at all points on the grid {\it simultaneously},
by looking, for each time $ t $, at a  time-$t$ snapshot of {\it all} Monte Carlo paths. Because we are in a setting of {\it batch-mode} Reinforcement Learning, the data are readily available in in form of a historical dataset.
  
\subsection{Batch Q-Learning: Fitted Q Iteration}

We use a most popular extension of 
Q-Learning to a batch RL setting called Fitted Q Iteration (FQI) \cite{Ernst, Murphy}.  The model formulation is now back to a general continuous-state space case, as in a Monte Carlo setting used in this paper, the only difference between a continuous-state and a discrete-state specification amounts to the choice of basis functions to use.



Assume that a dataset in the form (\ref{F_t_RL}) is available either 
as a simulated data, or as a real historical data. A starting point of the Fitted Q Iteration (FQI) \cite{Ernst, Murphy}  method 
is a choice of a parametric family of models for quantities of interest, namely optimal action and optimal action-value function. We use linear architectures where
functions sought are {\it linear} in adjustable parameters that are next optimized to find the optimal action and action-value function. 

We use the same set of basis functions  $ \{ \Phi_n(x) \} $ as we used above in Sect.~\ref{sect:DP}.
As the optimal Q-function $ Q_t^{\star} \left(X_t, a_t \right) $ is a quadratic function of $ a_t $, we can represent 
it as an expansion  
 in basis functions, with time-dependent coefficients parametrized by a matrix $ {\bf W}_t $:
\bea
\label{Q_any_a}
Q_t^{\star} \left(X_t, a_t \right) &=&
\left( 1, a_t, \frac{1}{2} a_t^2 \right) 
 \, \left( \begin{array}{cccc}
W_{11}(t)  & W_{12}(t) & \cdots & W_{1M}(t)   \\
W_{21}(t)  & W_{22}(t) & \cdots  & W_{2M}(t) \\  	
W_{31} (t) & W_{32}(t) & \cdots & W_{3M} (t)   \\
\end{array} \right)  
\left( \begin{array}{c}
\Phi_1(X_t)  \\
\vdots \\
\Phi_M(X_t)   \\
\end{array} \right) 
 \nonumber \\ 
&\equiv & {\bf A}_t^T   {\bf W}_t  {\bf \Phi}(X_t)
\equiv   {\bf A}_t^T  \, {\bf U}_{W} (t,X_t)  
\eea 
Eq.(\ref{Q_any_a}) is further re-arranged to convert it into a product of a parameter vector and a vector that 
depends on both the state and the action:    
\bea
\label{rearrange}
 Q_t^{\star} \left(X_t, a_t \right)  &=& {\bf A}_t^T   {\bf W}_t  {\bf \Phi}(X) = 
 \sum_{i=1}^{3} \sum_{j=1}^{M} \left( {\bf W}_t \odot  \left( {\bf A}_t  \otimes  {\bf \Phi}^T(X)  
 \right) \right)_{ij}  \nonumber \\
 &=& \vec{{\bf W}}_t \cdot vec 
 \left( {\bf A}_t  \otimes
 {\bf \Phi}^T(X) \right) \equiv  \vec{{\bf W}}_t \vec{{\bf \Psi}} \left(X_t,a_t \right)
 \eea
  
Here $ \odot $ stands for an element-wise (Hadamard) product of two matrices. The vector of time-dependent parameters 
$  \vec{{\bf W}_t}  $ is obtained by concatenating columns of matrix $ \bf{W}_t $, and similarly, 
$ \vec{{\bf \Psi}} \left(X_t,a_t \right) = 
 vec \left( {\bf A}_t  \otimes {\bf \Phi}^T(X) \right) $ stands for 
a vector obtained by concatenating columns of the outer product of vectors $ {\bf A}_t $ and $ {\bf \Phi}(X) $.
 
Coefficients  $  \bf{W}_t  $ can now be computed recursively backward in time for $ t = T-1, \ldots, 0 $.
To this end, the one-step Bellman optimality equation (\ref{Bellman_Q}) 
is interpreted as regression of the form
\beq
\label{Q_star_rec_4}
 R_t \left(X_t, a_t, X_{t+1} \right)   + \gamma \max_{a_{t+1} \in \mathcal{A}}  
  Q_{t+1}^{\star} \left( X_{t+1}, a_{t+1} \right)  =  \vec{{\bf W}}_t \vec{{\bf \Psi}} \left(X_t,a_t \right) + \varepsilon_t
\eeq
where $ \varepsilon_t $ is a random noise at time $ t $ with mean zero. 
Eqs. (\ref{Q_star_rec_4}) and (\ref{Bellman_Q}) are equivalent in expectations, as 
taking the expectation of both sides of (\ref{Q_star_rec_4}), we recover (\ref{Bellman_Q}) with 
function approximation (\ref{Q_any_a}) used for the optimal Q-function $ Q_t^{\star} \left(x, a \right) $. 

Coefficients $  \bf{W}_t  $ are therefore found by solving the following least-square optimization problem:
\beq
\label{W_problem}
\mathcal{L}_t \left(  {\bf W}_t \right) = \sum_{k=1}^{N_{MC}} \left(  R_t \left(X_t, a_t, X_{t+1} \right)   + \gamma \max_{a_{t+1} \in \mathcal{A}}  
  Q_{t+1}^{\star} \left( X_{t+1}, a_{t+1} \right)  -  \vec{{\bf W}}_t \vec{{\bf \Psi}} \left(X_t,a_t \right) \right)^2
\eeq 
Note that this relation holds for a general {\it off-model}, {\it off-policy} setting of the Fitted Q Iteration method of RL. 

Performing minimization, we obtain
\beq
\label{W_opt_vec}
 \vec{\bf{W}}_{t}^{\star} =  {\bf S}_t^{-1} {\bf M}_t
\eeq
where
\bea
\label{SM}
S_{nm}^{(t)} &=& \sum_{k=1}^{N_{MC}} \Psi_n \left( X_t^k, a_t^k \right)  \Psi_m \left( X_t^k, a_t^k \right)
\nonumber \\
M_{n}^{(t)} &=& \sum_{k=1}^{N_{MC}} \Psi_n  \left( X_t^k, a_t^k \right)  
 \left(   R_t \left(X_t^k, a_t^k, X_{t+1}^k \right)   + \gamma \max_{a_{t+1} \in \mathcal{A}}  
  Q_{t+1}^{\star} \left( X_{t+1}^k, a_{t+1} \right)   \right)
\eea
To perform the maximization step in the second equation in (\ref{SM}) analytically, note that 
because coefficients ${\bf W}_{t+1} $ and hence vectors $  {\bf U}_{W} (t+1, X_{t+1})  \equiv  {\bf W}_{t+1} {\bf \Phi}(X_{t+1})
$ (see Eq.(\ref{Q_any_a})) are known from the previous step, we have
\beq
\label{max_Q}
Q_{t+1}^{\star} \left( X_{t+1}, a_{t+1}^{\star}  \right) =   \mathbf U_W^{\left(0\right)} \left(t+1,X_{t+1} \right) +
a_{t+1}^{\star} \mathbf U_W^{\left(1\right)} \left(t+1,X_{t+1} \right) + 
 \frac{\left( a_{t+1}^{\star} \right)^2 }{2} \mathbf U_W^{\left(2\right)} \left(t+1,X_{t+1} \right) 
\eeq
It is important to stress here that while this is a quadratic expression in $ a_{t+1}^{\star} $, it would be {\it wrong} to use a point of its maximum as a function of  $ a_{t+1}^{\star} $ as such optimal value in Eq.(\ref{max_Q}).  
This would amount to using the same dataset to estimate both the optimal action and the optimal Q-function, leading to an overestimation of  $ Q_{t+1}^{\star} \left( X_{t+1}, a_{t+1}^{\star}  \right)  $ in Eq.(\ref{SM}), due to Jensen's inequality and convexity of the $ \max(\cdot ) $ function.
 The correct way to use Eq.(\ref{max_Q}) is to plug there a value of $ a_{t+1}^{\star} $ computed using the analytical solution 
 Eq.(\ref{a_star_t}) (implemented in the sample-based approach in Eq.(\ref{phi_nt_vec})), applied at the previous time step. Due to availability of the analytical optimal action  (\ref{a_star_t}),
 a potential overestimation problem, a classical problem of Q-Learning that is sometimes addressed using such methods as Double Q-Learning 
 \cite{Double-Q}, is avoided in the QLBS model, leading to numerically stable results.

\subsection{Discussion}
\label{sect:Discussion_FQI}

Equation (\ref{W_opt_vec}) gives the solution for the QLBS model in a model-free, data-driven way that bypasses the problem of defining the dynamics of the 
world.  
Computationally, our model needs only little linear algebra (linear regression), along with a set of basis functions. 
Linear regressions in this model take actual {\it trading data} from a dynamic portfolio of a stock and cash, and convert it {\it directly} into the optimal price and optimal hedge for the option.
Parameters found by these regressions (weights of basis functions) are parameters that define the optimal price and optimal hedge strategy themselves,
there are {\it no} other tunable parameters in our approach.
Unlike more traditional option pricing models, our approach obliviates any need for any additional model estimation/calibration.


Under the {\it only} modeling assumption that local rewards (negative losses) are quadratic a-l{\'a} Markowitz 
\cite{Markowitz},  Eq.(\ref{W_opt_vec}) provides a model-free data-driven solution to the option pricing problem. It only needs training data from an option trading desk in the form of tuples (\ref{F_t_RL}), plus basic linear algebra that can be implemented using GPU architectures, if a computational speed becomes critical. No volatility surface models or jump-diffusion models with integro-differential equations for option pricing are ever needed, and not even canonical $ N(d_1) $ and $ N(d_2) $ of the BSM model (see e.g. \cite{Wilmott}) are nowhere to be seen.
As long as rewards are quadratic, the Q-Learner goes {\it model-free} and produces the model-independent result (\ref{W_opt_vec}) that defines both the optimal hedge and optimal price.   


Interestingly, actions in training data can be completely random, and the Q-Learner will still learn the right price and hedge of an option if there is enough {\it market data}. This is guaranteed by the property of Q-Learning and Fitted Q Iteration of being {\it 
off-policy} algorithms. This creates an ability to create artificial trading histories by combining real market data (stock prices) with 
some pre-defined trading strategy in stocks.  


\section{Possible numerical experiments}
\label{sect:QL_BSM_numeric}

The main result of this paper is given by by a simple recursive relation Eq.(\ref{W_opt_vec}), implementable 
via linear regression, that provides a way to price and hedge a stock option by learning {\it directly from the past trading data}.
Our model is backed by the celebrated Q-Learning method of Watkins \cite{Watkins_1989, Watkins} and the Fitted Q Iteration method of Ernst {\it et. al.} \cite{Ernst} and Murphy \cite{Murphy}. 

As the latter methods are {\it model-free} and converging, and the explicit {\it model-free} solution of the model is provided by Eq.(\ref{W_opt_vec}), it was the author's feeling that this also makes {\it him} free to concentrate in this paper on the mathematical solution of the model, and  leave numerical examples and simple extensions for a companion paper \cite{QLBS_NuQLear}.

  
\section{Summary}
\label{sect:Summary}

In this paper we presented the QLBS model - a model for derivatives prices that is rooted in Reinforcement Learning.
It was designed with an idea of building a model for derivatives pricing that would implement the principle of hedging first and pricing second in a consistent way for a discrete-time version of the classical Black--Scholes(-Merton) model. 
By definition, an optimal action-value $Q$-function \cite{SB} does such {\it hedging and pricing by learning}, if we have an option pricing model where 
the option price is the negative of a $ Q$-function, and the option hedge is its second argument. Once we construct such a model, we can use Q-Learning, one of the most powerful algorithms of Reinforcement Learning, to learn both the optimal option price and hedge {\it directly from trading data} for a replicating portfolio of a stock and a bank cash account. Therefore, by 
stepping {\it aside} from the 
academic limit $ \Delta t \rightarrow 0 $, our model gains model independence, because it is solved by Q-Learning, and even  
gets somewhat simplified numerically, as an added bonus!

Author's original intention at the start of this project was to have a toy ``BSM-like" simulated financial environment in a simple 
discrete-time/discrete-space setting, that would be used for pedagogical purposes to explain simple RL algorithms such as Q-Learning. We hope that, at the very least, the QLBS model did achieve this goal, and therefore could potentially be  
used, along with its straightforward extensions, for benchmarking different RL algorithms in a simulated financial environment. Due to a high degree of tractability of our model (linear algebra plus Monte Carlo simulation) and its extensibility, both the discrete-space and continuous-space versions of the model can be used to benchmark different RL algorithms including policy gradient methods, actor-critic algorithms, Deep Reinforcement Learning, and so on - all with known optimal policies and action-value functions easily computable by means of DP. 

However, as the initial idea of this project, to have ``something like the BSM" for RL in Finance, has been progressing, different elements of this scheme started to come together as a {\it real} financial model, especially after recognizing the fact that Q-Learning gives, by its definition, {\it both} the price and the hedge in one easily computable formula, if only the option price is defined properly. Given both the simplicity and generality of 
this emerging model, it started to appear that the Q-Learner was there all the time, but just on the side, keeping its distance 
 $ \Delta t > 0 $ and quietly waiting to get noticed, while the vast majority\footnote{Except the literature on hedging and pricing in incomplete markets.} 
 of Mathematical Finance literature focused on puzzles of  continuous-time $ \Delta t \rightarrow 0 $ worlds (both within the BSM model itself and its many ``risk-neutral" extensions). 
The model presented in this paper is in some way even simpler than the continuous-time BSM model which involves non-elementary functions such as cumulative normal distributions of composite arguments, which correspond to infinite series in terms of elementary functions.
Our model involves only Linear Algebra and finite sums, and both hedge and price are contained in {\it one} formula, not two as in the BSM model, though the formula itself is not given in closed form. 

If the famous Black-Scholes {\it formula} for the option price was somehow {\it not} known till today, it could be re-discovered by the QLBS model, by analyzing data collected with purely random strategies, combined with a Deep Neural Network to approximate the value function, in a limit of very small time steps. This is because Q-Learning is an {\it off-policy} method that can learn even from data with purely random actions.  The only question is how much data and how many GPUs or TPUs are needed.
On the other hand, {\it no knowledge} of Partial Differential Equations would be required with such approach! Experiments of this kind will possibly be presented elsewhere. 
      
Besides mathematical simplicity, and probably more importantly,  because Q-Learning is a model-free method, our approach can be used to price and hedge options {\it directly from data, and without any model at all}, going ``the RL way" by extending this framework to a multi-dimensional case, and using more advanced versions of Q-Learning.  

On the financial side, our model demonstrates that one very classical financial model (the continuous-time BSM model) can be obtained from another classical financial model (the Markowitz portfolio theory), if applied in a special dynamic setting of a BSM replicating portfolio under BSM lognormal dynamics, taken to the limit  $ \Delta t \rightarrow 0 $. As is known since 
F{\"o}llmer and Schweizer \cite{FS, Schweizer}, this limit washes out any {\it risk}. But {\it markets only exist when there is risk to trade}.
Therefore, the very last step of going  to the limit  $ \Delta t \rightarrow 0 $ in the classical BSM model, while making perfect sense 
{\it mathematically}, makes no sense {\it financially}. While this latter fact is {\it also} well known in the literature, our model is the first one that not only provides an easy {\it consistent} pricing and hedging alternative for a discrete-time case $ \Delta t > 0 $, but also does it in a data-driven and model independent way, effectively reducing {\it all} of option pricing to Reinforcement Learning. 

The QLBS model provides a {\it simultaneous and consistent} model-independent and RL-based hedging and pricing of a European option (the method is straightforward to extend to other options e.g. American, Asian, basket, etc.). This is unlike many other incomplete market models where a link between pricing and hedging is more {\it ad hoc}. The QLBS model extends the BSM model by adding three more quite intuitive parameters: the stock drift $ \mu $, the time step $ \Delta t $, and the Markowitz risk-aversion parameter 
$ \lambda$\footnote{Note that in the natural sciences, models with four or five parameters are often able to capture even very complex dynamic systems 
\cite{von_Neumann}.}.

The QLBS approach can be extended to a option portfolios along the lines outlined in \cite{QLBS_NuQLear}, thus offering a principled approach to the infamous volatility smile problem of the classical BSM model (see e.g. \cite{Wilmott}).  Due to Q-Learning and reliance on data instead of a model, the volatility smile problem simply {\it does not exist} for the Q-Learner - it is just a name that the humans gave to a specific pattern in data.

Unlike the classical BSM model rooted in Ito's calculus, our QLBS model is rooted in Dynamic Programming and Reinforcement Learning, and capitalizes on convergence results for $Q$-learning to establish its own convergence to the classical BSM model in the academic limit $ \Delta t \rightarrow 0 $ (if the world is lognormal), while producing corrections outside this limit, by finding them {\it directly from trading data} using $Q$-Learning algorithms.  Like the Cheshire Cat from Carrol's 
 ``Alice's Adventures in Wonderland", 
the risk-averse Q-Learner of our model disappears in the strict limit $ \Delta t \rightarrow 0 $, leaving only its grin (i.e. the BS price and delta of the option) behind. As all risk is gone in this limit, and markets are fake, its services are no longer needed, and it leaves the degenerate $ \Delta t \rightarrow 0 $ BSM world, in its quest for more {\it purposeful}, rational, BSM-like or otherwise, {\it risky} worlds. We know that it will be just fine out there, because it is {\it model-free}!


\def\thesection{A}	
\setcounter{equation}{0}
\def\theequation{\thesection.\arabic{equation}}

\section*{Appendix A: Indifference Hedging and Pricing}

The framework presented in Sect.~\ref{sect:DT_BSM} largely follows the Hedged Monte 
Carlo (HMC) method of Potters and Bouchaud  \cite{PB}, with modifications suggested by Grau \cite{Grau}.
This approach is based on a quadratic loss (utility) function.
Here we outline an alternative approach to the pricing and hedging under
market incompleteness. 

We use the expected utility maximization approach with 
a specific choice of the exponential utility function
$ U(X) = - \exp( - \gamma X) $. As will be shown below,
as long as the risk aversion parameter $ \gamma $ is small, the only optimization that is 
needed in this approach is convex optimization arising at the stage of finding 
the Minimum Entropy Martingale Measure (MEMM) $ \mathbb{Q} $.

The latter optimization can still be computationally costly as it should be done at 
each node at each time slice. However, if the model for the underlying 
is simple enough (e.g. 
an arithmetic or geometric Brownian motion), the transition from the real measure 
$ \mathbb{P} $ to a 
risk-neutral martingale measure $ \mathbb{Q} $ is known analytically in closed form.
In this case, assuming that we have already computed the risk-neutral measure and moreover
we simulate from this measure, no additional numerical optimization is needed in the 
formalism to be presented below, as long as the risk aversion parameter $ \gamma $ is small.
Instead, we use a low-order perturbative expansion in powers of 
$ \gamma $ to approximate all quantities 
of interest (i.e. hedges and indifference prices of options)\footnote{
Note that the usual Minimum Entropy (MinEnt) model for option prices is recovered in this
approach as the zero-order term of this expansion. In other words, MinEnt corresponds
to a risk-neutral limit $ \gamma \rightarrow 0 $ of the result of expected utility maximization 
for the exponential utility.}.   
In this case, the whole calculation to the zero order in $ \gamma $
proceeds as in the quadratic risk formulation, the 
difference being that we now compute with the MEMM $ \mathbb{Q} $ rather 
than a real-world measure $ \mathbb{P} $. 
In addition, to calculate first-order corrections $ O(\gamma) $, all we need is to compute
two additional $ \mathbb{Q} $-expectations   
per node\footnote{
If desired, this scheme can be easily extended to an arbitrary finite order in $ \gamma $.
The cost of calculating each additional correction in expansion in powers of 
$ \gamma $ is again two additional $ \mathbb{Q} $-expectations per node.}. 

\subsection*{Exponential utility and duality}

Here we provide a brief overview of a related method developed by Lim \cite{Lim} in the setting of 
optimal option hedging and pricing  with an exponential utility. For details, proofs, and references to  
original publications, please see Ref.~\cite{Lim}.

To price and hedge a contingent claim with payoff $ B $, we consider two optimal 
investment problems, with and without the derivative $ B $:
\bea
\label{optInvest}
V(x,t) &=& \sup_{u \in \mathcal{A} } \mathbb{E}^{P} \left[
\left. - e^{ - \gamma X^{x,u}(T)} \right| X^{x,u}(t) = x, \mathcal{F}_t \right]
\nonumber \\
V^B(x,t) &=& \sup_{u \in \mathcal{A} } \mathbb{E}^{P} \left[
\left. - e^{ - \gamma (X^{x,u}(T) - B)} \right| X^{x,u}(t) = x, \mathcal{F}_t \right]
\eea
Here $ u = u_t  $ is the investment strategy and $ x $ is the initial wealth at time 
$ t $. We concentrate on the option price for the writer of the option.

The indifference price $ h_t(B) $ 
for claim $B $ is defined as an extra wealth that makes the option writer
indifferent between writing and not writing the option under the optimal investment strategy.
It is found as a solution to the equation
\beq
\label{indifferencePrice}
V(x,t) = V^B \left( x + h_t(B), t \right)
\eeq
The key relation is the duality formula that replaces the 
real-measure maximization wrt strategies $ u \in \mathcal{A} $ 
by maximization wrt equivalent martingale measures $ \mathcal{M}_e $ of $ \mathbb{P} $:
\beq
\label{duality}
\sup_{u \in \mathcal{A} } \mathbb{E}^{P} 
\left[ \left. - e^{ - \gamma (X^{x,u}(T) - B)} \right| \mathcal{F}_t \right]  
 = - e^{ - \gamma x} \exp \left(  \sup_{M \in \mathcal{M}_e}
\left\{  - H_t(M|P) + \gamma \mathbb{E}^M[B | \mathcal{F}_t] \right\} \right)
\eeq
where $ H_t(M|P) $ stands for the conditional entropy of measures $ M $ and $ P $.
Using (\ref{indifferencePrice}) and (\ref{duality}), the indifference price reads
\bea
\label{indiffPriceFinal}
h_t(B) &=& \frac{1}{\gamma} \left[  \sup_{M \in \mathcal{M}_e} \left\{
- H_t(M|P) + \mathbb{E}^M \left[ \gamma B| \mathcal{F}_t \right] \right\}
- \sup_{M \in \mathcal{M}_e} \left\{ - H_t(M|P) \right\} \right] \nonumber \\
& \equiv & \frac{1}{\gamma} \left[ v_t^{P,\gamma}(B) - v_t^{P}(0) \right]
\eea
The indifference pricing therefore reduces to two functional optimizations for two
terms in (\ref{indiffPriceFinal}). These optimizations are done using 
the Lagrange multiplier method. Let $ S_i = (S_{i1},\ldots, S_{id} ) $ be the 
node for the underlying at time $ t $ (here $ d  $ is the number of components), 
$ \mathcal{I}_t $ be the set of all nodes at can be reached starting from node 
$ i $ at $t $, and $ \Delta S_{ij} $ is a vector of changes of the underlying from
node $ i $ at $ t $ to node $ j $ at time $ t + 1 $.
We have
\bea
\label{minmax}
v^{P,\gamma}(t,i,B) &=& \min_{\mu,\lambda} \max_{m} \left\{ - \sum_{j \in \mathcal{I}_t}
m_j \left[ \log \frac{m_j}{p_j} - \gamma b_j \right] + \mu \left( 
\sum_{j \in \mathcal{I}_t} m_j - 1 \right) + \lambda_B^T 
\sum_{j \in \mathcal{I}_t} m_j 
\Delta S_{ij} \right\} \nonumber \\
&=& \min_{\lambda} \left[ \log Z_{ti}(\lambda_B) \right] 
\eea
where $ \lambda_B = (\lambda_B^{(1)}, \ldots, \lambda_B^{(d)}) $ is a vector of 
Lagrange multipliers, and
\beq
\label{Z}
Z_{ti} = \sum_{j} p_j e^{ \gamma b_j + \lambda_B^{T} \Delta S_{ij}}
\eeq
The remaining minimization wrt Lagrange multipliers $ \lambda_B $ amounts to convex
optimization in $ d $ dimensions. The second functional $ v_t^P(0) $ is computed similarly
with the substitution $ b_j \rightarrow 0 $, $ \lambda_B \rightarrow \lambda_0 $.
 
Note that Eqs.(\ref{Z}) and (\ref{indiffPriceFinal}) indicate that the indifference price
is non-linear in the payoff. This is in contrast to the linear pricing rule 
of the Minimum Entropy method. However, it is easy to check that the latter is 
recovered from the indifference 
pricing method in the limit $ \gamma \rightarrow 0 $. This can be seen by expanding 
$ v_t^{P,\gamma} $ to the first order around $ v_t^{P,0} $ and then taking the limit 
$ \gamma \rightarrow 0 $ in (\ref{indiffPriceFinal}), see the next section.
 
The optimal hedge ratios $ u_k $ ($ k = 1,\ldots, d $) are obtained as a rescaled difference
of two optimal Lagrange multipliers calculated with and without the option:
\beq
\label{optHedge}
u_k =  - \frac{1}{\gamma} \left( \lambda_{B,k} - \lambda_{0,k} \right)
\eeq

The above formulae refer to a single-period setting, and involve two Lagrangian optimizations
with two set of Lagrange multipliers. A multi-period recursive 
version involving only the MEMM $ \mathbb{Q} $ was worked out by Lim \cite{Lim}:
\beq
\label{multiperiodPrice}
h_{t} = \frac{1}{\gamma} \log
\mathbb{E}_{(t)}^{Q} \left[ e^{ \gamma (h_{t+1} - u_{t+1}^T \Delta S_{t} )} \right] = 
u_{t+1}^T S_t + \frac{1}{\gamma} \log
\mathbb{E}_{(t)}^{Q} \left[ e^{ \gamma (h_{t+1} - u_{t+1}^T S_{t+1} )} \right]
\eeq
for $ t = 0,\ldots, T-1 $ and $ h_{T,j} = B_j $, where the optimal hedge is defined as follows
\beq
\label{optHedge2}
u_{t+1} = \arg\min_{u} \mathbb{E}_{(t)}^{Q} \left[e^{ \gamma (h_{t+1} - 
u^T \Delta S_{t} )} \right] 
\eeq	

 \subsection*{Expansions for small $ \gamma $}

While equations (\ref{multiperiodPrice}) and (\ref{optHedge2}) are valid for arbitrary 
values $ \gamma \geq 0 $, in practice their use can be costly for the calculation 
of optimal hedges, as the latter amounts to a transcendent equation that should be solved 
numerically at each node. Instead, we invoke a low-order expansion in $ \gamma $ in order
to calculate approximate optimal hedges and option prices.

We start with the optimal hedge formula (\ref{optHedge2}). We look for the optimal hedge 
ratio in terms of an expansion in powers of $ \gamma $:
\beq
\label{hedgeExpansion}
u_{t+1} = u_{t+1}^{(0)} + \gamma u_{t+1}^{(1)} 
+ \gamma^2 u_{t+1}^{(2)} + \ldots
\eeq
Plugging this into (\ref{optHedge2}) and equating the like powers of $ \gamma $, we find
\bea
u_{t+1}^{(0)} &=& \frac{ Cov_{t}^{Q} \left[ h_{t+1}, \Delta S_{t} \right]}{
Var_{t}^{Q} \left[ \Delta S_t \right]}  \\
u_{t+1}^{(1)} &=& \frac{1}{2} 
\frac{ \mathbb{E}_t^Q \left[  \left(h_{t+1} - u_{t+1}^{(0)T} \Delta S_t \right)^2 
\Delta S_t \right]}{
Var_t^Q \left[ \Delta S_t \right] } \nonumber 
\eea
If desired, this expansion can be continued. 

Next we turn to the pricing formula. It is convenient to introduce the following 
short notation:
\beq
\label{shortNot}
\tilde{h}_{t+1} = h_{t+1} -  u_{t+1}^{(0) T} \Delta S_t  \; \; , \; \;
\bar{\tilde{h}}_{t+1} = \mathbb{E}_t^Q \left[\tilde{h}_{t+1} \right]
\eeq
Expanding the second term in (\ref{multiperiodPrice}) in powers of $ \gamma $, we find
\bea
\label{mpPriceExpansion}
h_{t} &=& \mathbb{E}_t^Q \left[ \tilde{h}_{t+1} \right] + \frac{1}{2} \gamma Var_t^Q
\left( \tilde{h}_{t+1} \right) + \frac{1}{3!} \gamma^2 \mathbb{E}_t^Q \left[ 
\left(\tilde{h}_{t+1} - \bar{\tilde{h}}_{t+1} \right)^3 \right] + \ldots \nonumber \\  
&=& \mathbb{E}_t^Q \left[ h_{t+1} \right] 
+ \frac{1}{2} \gamma Var_t^Q
\left( \tilde{h}_{t+1} \right) 
+ \frac{1}{3!} \gamma^2 \mathbb{E}_t^Q \left[ 
\left(\tilde{h}_{t+1} - \bar{\tilde{h}}_{t+1} \right)^3 \right]   + \ldots
\eea
where in the second line we used the definition (\ref{shortNot}) and the fact that the 
Q-expectation of $ \Delta S_t $ is zero. 
 
Eq.(\ref{mpPriceExpansion}) is an expansion in moments of the slippage 
distribution. Note that the first two terms 
of this formula are very similar to the recursive pricing formula used in the local quadratic risk formulation, except 
that the conditional expectation is calculated under the risk-neutral
measure $ \mathbb{Q} $ rather than the physical measure $ \mathbb{P} $.
Also note that the resulting expression is non-negative provided contributions 
from higher moments (the third and above) are small. Therefore, this approach solves the 
problem with negative option prices in the quadratic risk formulation\footnote{If a truncated 
expansion produces a negative option price due to, say, a 
negative third moment, this can always be fixed by either adding more moments to the expansion
(\ref{mpPriceExpansion}), or by doing a full-blown numerical hedge optimization, where 
option prices are guaranteed to stay positive.}.

Next consider correction terms. The leading correction is proportional to the 
variance of the slippage distribution, with the coefficient fixed by the risk aversion parameter
$ \gamma $. The next correction is proportional to the third moment, etc. Therefore, this 
approach gives support to the intuitive idea of Potters and Bouchaud \cite{PB} that a risk 
premium in the option price should be driven by the second moment of the P\&L distribution.
It is exactly what happens in our approach, though the measure is fixed to be $ \mathbb{Q} $ 
rather than $ \mathbb{P}$, and the coefficient is fixed in terms of $ \gamma $. 

\def\thesection{B}	
\setcounter{equation}{0}
\def\theequation{\thesection.\arabic{equation}}

\section*{Appendix B: Least Square Monte Carlo (LSM) for American Options}

While the objective of the American Monte Carlo method of Longstaff and Schwartz \cite{LS} is altogether different from
the problem addressed in this paper (a risk-neutral valuation of an American option vs a real-measure discrete-time 
hedging/pricing of a European option), a {\it mathematical} setting is similar. Both problems look for an optimal strategy and solve this problem by a backward recursion in a combination with a forward Monte Carlo simulation.
Here we provide a brief outline of their method.

The main idea of the LSM approach of Longstaff and Schwartz \cite{LS} is to treat the backward-looking 
stage of the security evaluation as a regression problem
formulated in a forward-looking manner which is more suited for a Monte Carlo (MC) 
setting.
The starting point is the (backward-looking) Bellman equation, the most fundamental equation of the stochastic optimal control (otherwise known as stochastic optimization). 
For an American option on a financial underlying, control variable takes a simple 
binary form (``exercise" or ``not exercise"). 
The Bellman equation for this particular case produces 
 the continuation value $ C_t(S_t) $ at time $ t $ as a function
of the current underlying value $ S_t $:
\beq
\label{Bellman}
C_t(S_t) = \mathbb{E} \left[ \left. e^{- r \Delta t} \max \left( h_{t + \Delta t}(S_{t + \Delta t}), 
C_{t + \Delta t}(S_{t + \Delta t}) \right) \right| \mathcal{F}_t \right]
\eeq
Here $ h_{\tau}(S_\tau) $ is the option payoff at time $ \tau $. For example, for an American
put option $ h_{\tau}(S_\tau) = \left( K - S_{\tau} \right)^{+}  $.
 
 Note that for American options, the continuation value should be estimated 
 as a {\it function} $ C_t (x) $ of the value $ x = X_t $, as long we we want to know whether it is larger
 or smaller than the intrinsic value $ H(X_t) $ for a particular {\it realization} $ X_t = x $ 
(obtained e.g. with Monte Carlo simulation) of the process $ X_t $ at time $ t $.
The problem is, of course, that each Monte Carlo path has exactly {\it one} value 
of $ X_t $ at time $ t $. One way to estimate a function $ C_t(S_t)  $ is 
to use all Monte Carlo paths, i.e. use the {\it cross-sectional information}.
To this end, the one-step Bellman equation (\ref{Bellman}) 
is interpreted as  regression of the form
\beq
\label{reg1}
\max \left( h_{t + \Delta t}(S_{t + \Delta t}), 
C_{t + \Delta t}(S_{t + \Delta t}) \right)  = e^{r \Delta t} C_t(S_t) + \varepsilon_t(S_t)
\eeq
where $ \varepsilon_t(S_t) $ is a random noise at time $ t $ with mean zero,
which may in general depend on the underlying value $ S_t $ at that time.
Clearly (\ref{reg1}) and (\ref{Bellman}) are equivalent in expectations, as 
taking the expectation of both sides of (\ref{reg1}), we recover (\ref{Bellman}).
Next the unknown function $ C_t(S_t) $ is expanded in a set of basis functions:
\beq
\label{basis}
C_t(x) = \sum_{n} a_n(t) \phi_n(x) 
\eeq
for some particular choice of the basis
$ \{ \phi_n(x) \} $, and the coefficients $ a_n(t) $ are then calculated using the 
least-squared regression of $ \max \left( h_{t + \Delta t}(S_{t + \Delta t}), 
C_{t + \Delta t}(S_{t + \Delta t}) \right) $ on the value $ S_t $ of the underlying at time $ t $ 
across all Monte Carlo paths.


\begin{thebibliography}{99}
\bibitem{von_Neumann} ``With four parameters I can fit an elephant, and with five I can make him wiggle his trunk." (John von Neumann).
\end{thebibliography}
\end{document}